\documentclass[twocolumn,showpacs]{revtex4}
\usepackage{epsfig}
\makeatletter
\newcommand{\ber}{\begin{eqnarray}}
\newcommand{\eer}{\end{eqnarray}}
\newcommand{\pp}{p_{\perp}}
\newcommand{\pt}{p_{\rm T}}
\newcommand{\lra}{\leftrightarrow}

\begin{document}
\title{A Multi-Phase Transport Model for Relativistic Heavy Ion Collisions}
\author{Zi-Wei Lin}
\affiliation{Physics Department, The Ohio State University,
Columbus, Ohio 43210\footnote{
Present address: 301 Sparkman Drive, VBRH E-39, The University of Alabama 
in Huntsville, Huntsville, AL 35899}}
\author{Che Ming Ko}
\affiliation{Cyclotron Institute and Physics Department, Texas A\&M University,
College Station, Texas 77843-3366}
\author{Bao-An Li}
\affiliation{Department of Chemistry and Physics, Arkansas State University,
State University, Arkansas 72467-0419}
\author{Bin Zhang}
\affiliation{Department of Chemistry and Physics, Arkansas State University,
State University, Arkansas 72467-0419}
\author{Subrata Pal}
\affiliation{Department of Physics, Michigan State University,
East Lansing, Michigan 48824}

\date{\today}

\begin{abstract}
We describe in detail how the different components of a multi-phase 
transport (AMPT) model, that uses the Heavy Ion Jet Interaction
Generator (HIJING) for generating the initial conditions, 
Zhang's Parton Cascade (ZPC) for modeling partonic scatterings, 
the Lund string fragmentation model or a quark coalescence model for
hadronization, and A Relativistic Transport (ART) model for treating
hadronic scatterings, are improved and combined to give a coherent 
description of the dynamics of relativistic heavy ion collisions. 
We also explain the way parameters in the model are determined, and 
discuss the sensitivity of predicted results to physical input in the
model. Comparisons of these results to experimental data, mainly from
heavy ion collisions at the Relativistic Heavy Ion Collider (RHIC),  
are then made in order to extract information on the properties of the
hot dense matter formed in these collisions. 

\end{abstract}

\maketitle

\section{introduction}
Colliding heavy ions at relativistic energies makes it possible to subject 
nuclear matter to the extreme condition of large compression, leading to
energy densities that can exceed that for producing a plasma
of deconfined quarks and gluons, that is believed to have existed
during the first microsecond after the Big Bang. Experiments at RHIC
at the Brookhaven National Laboratory with center-of-mass energy up to
$\sqrt{s_{NN}}=200$ GeV 
in Au+Au collisions thus provide the opportunity to study
the properties of this so-called quark-gluon plasma (QGP). At the future
Large Hadron Collider (LHC) at CERN, which will allow Pb+Pb collisions
at $\sqrt{s_{NN}}=5.5$ TeV, the produced quark-gluon plasma will have an even
higher temperature and a nearly vanishing net baryon chemical potential. 

Many observables have been measured at RHIC, such as the rapidity 
distributions of various particles and their transverse momentum
spectra up to very high transverse momentum, the centrality dependence
of these observables, the elliptic flows of various particles, as well as both
identical and non-identical two-particle correlations. To understand
these extensive experimental results, many theoretical models have
been introduced. It ranges from thermal models
\cite{Braun-Munzinger:1994xr,Braun-Munzinger:1999qy,Cleymans:1998fq,Becattini:2000jw}
based on the assumption of global thermal and chemical
equilibrium to hydrodynamic models 
\cite{Hung:1994eq,Rischke:1995ir,Rischke:1995mt,Kolb:2000fh,Huovinen:2001cy,Kolb:2001qz,Teaney:2001av}
based only on the assumption of local thermal equilibrium, 
and to transport models 
\cite{Sorge:vt,Pang:1992sk,Li:1995pr,Sa:1995fj,Sorge:1995dp,Jeon:1997bp,Zhang:1997ej,Bass:1998ca,Humanic:ji,Tai:1998hc,Kahana:1998wf,Molnar:2000jh,Kahana:ky,Li:2001xh,Sa:2001ma} 
that treat non-equilibrium dynamics explicitly. The thermal models
have been very successful in accounting for the yield of various
particles and their ratios, while the hydrodynamic models are
particularly useful for understanding the collective behavior of low
transverse momentum particles such as the elliptic flow 
\cite{Kolb:2000fh,Huovinen:2001cy,Kolb:2001qz,Teaney:2001av}. 
Since transport models treat chemical and thermal freeze-out dynamically, 
they are also natural and powerful tools for studying the Hanbury-Brown-Twiss
interferometry of hadrons. For hard processes that involve large momentum
transfer, approaches based on the perturbative quantum chromodynamics
(pQCD) using parton distribution functions in the colliding nuclei 
have been used \cite{pqcd,Wang:1996yf}. Also, the classical
Yang-Mills theory has been developed to address the evolution of parton 
distribution functions in nuclei at ultra-relativistic energies
\cite{McLerran:1993ni,McLerran:1993ka,Kovchegov:1997ke}
and used to study the hadron rapidity distribution and its centrality 
dependence at RHIC \cite{Kharzeev:2000ph,Kharzeev:2001gp,Kharzeev:2002ei}. 
These problems have also been studied in the pQCD based final-state 
saturation model \cite{Eskola:1999fc,Eskola:2000xq,Eskola:2002qz}. 

Although studies based on the pQCD \cite{Baier:2000sb} have shown that 
thermalization could be achieved in collisions of very large nuclei
and/or at extremely high energy, even though the strong coupling constant 
at the saturation scale is asymptotically small, the dense matter
created in heavy ion collisions at RHIC may, however, not achieve
full thermal or chemical equilibrium as a result of its finite volume
and energy. To address such non-equilibrium many-body dynamics, 
we have developed a multi-phase transport (AMPT) model that
includes both initial partonic and final hadronic interactions and the
transition between these two phases of matter
\cite{Zhang:1999mq,Zhang:2000bd,Zhang:2000nc,Lin:2001cx,Lin:2001yd,Pal:2001zw,Lin:2001zk,Zhang:2002ug,Pal:2002aw,Lin:2002gc,Lin:2003ah,Lin:2003iq}.  
The AMPT model is constructed to describe nuclear collisions ranging 
from $p+A$ to $A+A$ systems at center-of-mass energies 
from about $\sqrt{s_{NN}}=5$ GeV up to 5500 GeV at LHC, 
where strings and minijets dominate the initial energy production 
and effects from final-state interactions are important. 
For the initial conditions, the AMPT model uses the hard minijet
partons and soft strings from the HIJING model. The ZPC parton cascade
is then used to describe scatterings among partons, which is
followed by a hadronization process that is based on the Lund string 
fragmentation model or by a quark coalescence model. The latter is
introduced for an extended AMPT model with string melting in which
hadrons, that would have been produced from string fragmentation, are
converted instead to their valence quarks and antiquarks. 
Scatterings among the resulting hadrons are described by the ART
hadronic transport model. With parameters, such as those in the string
fragmentation, fixed by the experimental data from heavy ion
collisions at SPS, the AMPT model has been able to describe reasonably
many of the experimental observations at RHIC. 

In this paper, we give a detailed description of the different
components of the AMPT model, discuss the parameters in the model, 
show the sensitivity of its results to the input to the
model, and compare its predictions with experimental data. The paper
is organized as follows. In Section \ref{ampt}, we describe the different
components of the AMPT model: the HIJING model and string melting, the
ZPC model, the Lund string fragmentation model, and the quark
coalescence model used for the scenario of string melting, and the extended ART
model. Tests of the AMPT model against data from $pp$ and $p\bar p$
reactions are given in Section \ref{pp}. Results from the AMPT model
for heavy ion collisions at SPS energies are discussed in Section
\ref{sps} for hadron rapidity distributions and transverse momentum
spectra, baryon stopping, and antiproton production.
In Section \ref{rhic}, we show results at RHIC for hadron
rapidity distributions and transverse momentum spectra, particle ratios, 
baryon and antibaryon production, and the production of 
multistrange baryons as well as $J/\psi$. We further show results from
the AMPT model with string melting on hadron elliptic flows and
two-pion interferometry at RHIC. In Section \ref{lhc}, we present the 
predictions from AMPT for hadron rapidity and transverse momentum
distributions in Pb+Pb collisions at the LHC energy. Discussions 
on possible future improvements of the AMPT model are presented 
in Section \ref{discussions}, and a summary is finally given in 
Section \ref{summary}.

\section{The AMPT model}
\label{ampt}

The AMPT model consists of four main components: the initial conditions, 
partonic interactions, the conversion from the partonic to the hadronic 
matter, and hadronic interactions. The initial conditions, which include 
the spatial and momentum distributions of minijet partons and 
soft string excitations, are obtained from the HIJING model 
\cite{Wang:1990qp,Wang:1991ht,Wang:1991us,Gyulassy:ew}. 
Currently, the AMPT model uses the HIJING model version 1.383 
\cite{hj1383}, which does not include baryon junctions 
\cite{junctionpapers}. 
Scatterings among partons are modeled by Zhang's parton cascade (ZPC)
\cite{Zhang:1997ej}, which at present includes only two-body
scatterings with cross sections obtained from the pQCD with screening 
masses. In the default AMPT model 
\cite{Zhang:1999mq,Zhang:2000bd,Zhang:2000nc,Lin:2001cx,Lin:2001yd,Pal:2001zw,Zhang:2002ug,Pal:2002aw,Lin:2003ah},  
partons are recombined with their parent strings when they stop
interacting, and the resulting strings are converted to hadrons using
the Lund string fragmentation model
\cite{Andersson:1983jt,Andersson:ia,Sjostrand:1993yb}. 
In the AMPT model with string melting \cite{Lin:2001zk,Lin:2002gc,Lin:2003iq}, 
a quark coalescence model is used instead to combine partons into hadrons. 
The dynamics of the subsequent hadronic matter is described by a hadronic 
cascade, which is based on the ART model \cite{Li:1995pr,Li:2001xh} 
and extended to include additional reaction channels that are
important at high energies. These channels include the formation and
decay of $K^*$ resonance and antibaryon resonances, and
baryon-antibaryon production from mesons and their inverse reactions
of annihilation. Final results from the AMPT model are obtained after 
hadronic interactions are terminated at a cutoff time ($t_{cut}$) 
when observables under study are considered to be stable, i.e.,
when further hadronic interactions after $t_{cut}$ will not significantly
affect these observables. We note that two-body partonic scatterings
at all possible times have been included because the algorithm of ZPC, which 
propagates partons directly to the time when the next collision occurs, is 
fundamentally different from the fixed time step method used in the ART model.

\begin{figure}[ht]
\centerline{\epsfig{file=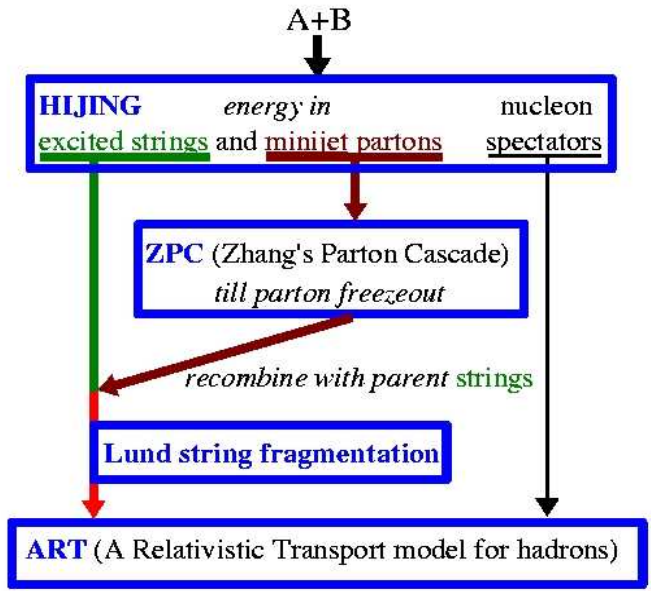,width=3.4in,height=3.4in,angle=270}}
\caption{(Color online) Illustration of the structure of the default
AMPT model.}
\label{flowchart1}
\end{figure}

\begin{figure}[ht]
\centerline{\epsfig{file=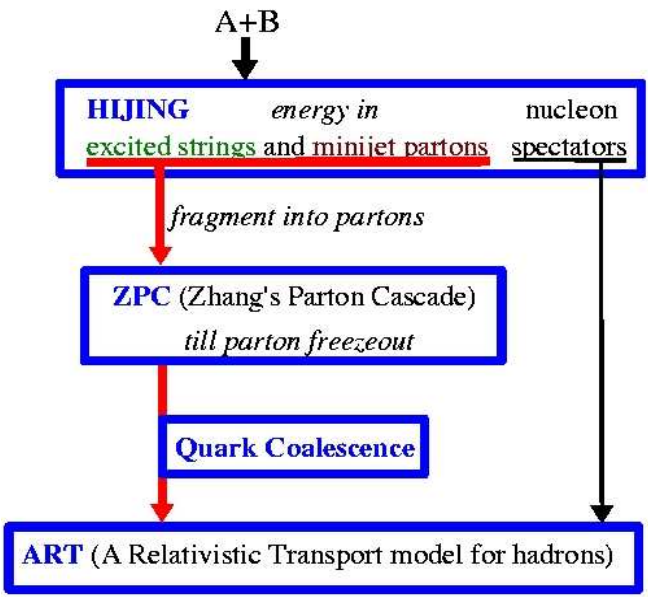,width=3.4in,height=3.4in,angle=270}}
\caption{(Color online) Illustration of the structure of the AMPT model 
with string melting.}
\label{flowchart2}
\end{figure}

In Figs.~\ref{flowchart1} and \ref{flowchart2}, we show, respectively,
the schematic structures of the default AMPT model
\cite{Zhang:1999mq,Zhang:2000bd,Zhang:2000nc,Lin:2001cx,Lin:2001yd,Pal:2001zw,Zhang:2002ug,Pal:2002aw}
and the AMPT model with string melting \cite{Lin:2001zk,Lin:2002gc,Lin:2003iq} 
described above.  The full source code of the AMPT model in the
Fortran 77 language and instructions for users are available online at
the OSCAR website \cite{oscar} and also at the EPAPS website \cite{epaps}. 
The default AMPT model is named as version $1.x$, and the AMPT model
with string melting is named as version $2.y$, where value of the
integer extension $x$ or $y$ increases whenever the source code is
modified. Current versions of the AMPT model is 1.11 for the default
model and 2.11 for the string melting model, respectively. In the
following, we explain in detail each of the above four components of the
AMPT model and the way they are combined to describe relativistic
heavy ion collisions.

\subsection{Initial conditions}\label{hijing}

\subsubsection{The default AMPT model}

In the default AMPT model, initial conditions for heavy ion
collisions at RHIC are obtained from the HIJING model
\cite{Wang:1990qp,Wang:1991ht,Wang:1991us,Gyulassy:ew}.  
In this model, the radial density profiles of the two colliding nuclei 
are taken to have Woods-Saxon shapes, and multiple scatterings among incoming 
nucleons are treated in the eikonal formalism. Particle production from two 
colliding nucleons is described in terms of a hard and a soft component. 
The hard component involves processes in which the momentum transfer
is larger than a cutoff momentum $p_0$ and is evaluated by the
pQCD using the parton distribution function in a nucleus. These hard
processes lead to the production of energetic minijet partons and 
are treated via the PYTHIA program. The soft component, on
the other hand, takes into account non-perturbative processes with
momentum transfer below $p_0$ and is modeled by the formation of
strings. The excited strings are assumed to decay independently
according to the Lund JETSET fragmentation model. 

From the $pp$ and $p\bar p$ total cross sections and the ratio of 
$\sigma_{\rm el}/\sigma_{\rm tot}$ in the energy range $20<\sqrt s<1800$ GeV, 
it has been found that the experimental data can be fitted with a 
nucleon-nucleon soft cross section $\sigma_s(s)=57$ mb at high energies 
and $p_0=2$ GeV/$c$ \cite{Wang:1990qp}. The independence of these two 
parameters on the colliding energy is due to the use of the 
Duke-Owens set 1 for the parton distribution function \cite{Duke:1983gd} 
in the nucleon. With different parton distribution functions, an 
energy-dependent $p_0$ may be needed to fit the same $pp$ and $p\bar p$ 
data \cite{Li:2001xa,ToporPop:2002gf}. 
We note that since the number of hard collisions in
an $A+A$ collision roughly scales as $A^{4/3}$ and grows fast with 
colliding energy while the number of strings roughly scales as $A$,
minijet production becomes more important when the energy of heavy 
ion collisions increases \cite{Wang:1990qp,Wang:2000bf}.

Because of nuclear shadowing, both quark \cite{quarkshadpapers} 
and gluon \cite{gluonshadpapers} 
distribution functions in nuclei are different from the simple superposition 
of their distributions in a nucleon. 
This effect has been included in the HIJING model via the following 
impact-parameter-dependent but $Q^2$(and flavor)-independent parameterization 
\cite{Wang:1991ht}: 
\ber
R_A(x,r) & \equiv & \frac {f_a^A (x,Q^2,r)}{A f_a^N (x,Q^2)} \nonumber \\
&=&1+1.19 \ln^{1/6}\!A (x^3-1.2 x^2+0.21 x ) \nonumber \\
&-& \! \left [\alpha_A(r) \!-\! \frac{1.08 (A^{1/3}-1) \sqrt x}{\ln (A+1)} \right ] e^{-x^2/0.01}, 
\eer
where $x$ is the light-cone momentum fraction of parton $a$, 
and $f_a$ is the parton distribution function. 
The impact-parameter dependence of the nuclear shadowing effect
is controlled by
\ber
\alpha_A(r)=0.133 (A^{1/3}-1) \sqrt {1-r^2/R_A^2},
\eer
with $r$ denoting the transverse distance of an interacting nucleon 
from the center of the nucleus with radius $R_A=1.2 A^{1/3}$. 
Note that there is a modified HIJING model which uses 
a different parameterization for the nuclear shadowing that is also 
flavor-dependent \cite{Li:2001xa}.

To take into account the Lorentz boost effect, we have introduced 
a formation time for minijet partons that depends on their four momenta 
\cite{Gyulassy:1993hr}. Specifically, the formation time for each 
parton in the default AMPT model is taken to have a Lorentzian distribution
with a half width $t_f=E/m_T^2$, where $E$ and $m_T$ are the parton 
energy and transverse mass, respectively. Initial positions of 
formed minijet partons are calculated from those of their parent 
nucleons using straight-line trajectories. 

\subsubsection{The AMPT model with string melting}
\label{initial-sm}

\begin{figure}[ht]
\centerline{\epsfig{file=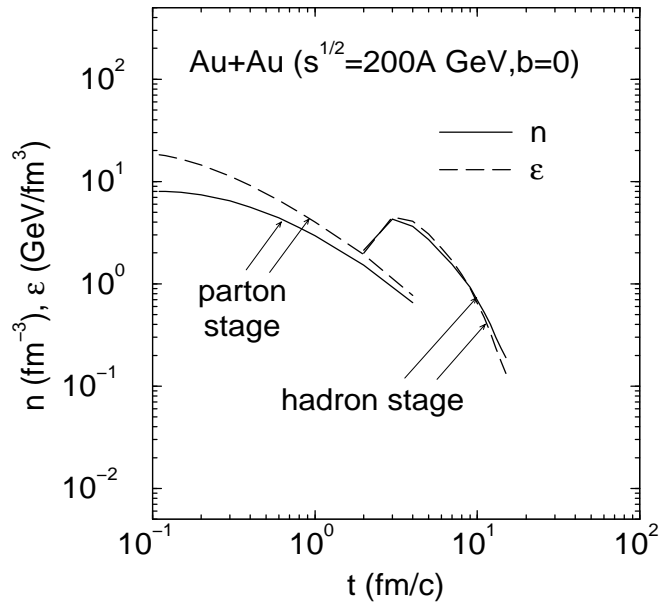,
width=3.4in,height=3.2in,angle=0}}
\caption{Energy and number densities of minijet partons and formed hadrons 
in the central cell as functions of time for central (b=0 fm) Au+Au 
collisions at $\sqrt {s_{NN}}=200$ GeV from the default AMPT model, 
where the energy stored in the excited strings is absent in the parton 
stage and is released only when hadrons are formed.}
\label{density}
\end{figure}

Although the partonic part in the default AMPT model includes only
minijets from the HIJING model, its energy density can be very high in 
heavy ion collisions at RHIC. As shown in Fig.~\ref{density} for
the time evolutions of the energy and number densities of 
partons and hadrons in the central cell of central ($b=0$ fm) 
Au+Au collisions at $\sqrt {s_{NN}}=200$ GeV in the center-of-mass 
frame, the partonic energy density during the first few fm/$c$'s of 
the collision is more than an order of magnitude higher 
than the critical energy density ($\sim 1$ GeV/fm$^3$) for the QCD phase 
transition, similar to that predicted by the high density QCD approach 
\cite{Kharzeev:2001ph}. The sharp increase in energy and number 
densities at about 3 fm/c is due to the exclusion of energies that 
are associated with the excited strings in the partonic stage. Keeping 
strings in the high energy density region \cite{Bjorken:1982qr} thus 
underestimates the partonic effect in these collisions. We note that
the central cell in the above calculation is chosen to have a transverse 
radius of 1 fm and a longitudinal dimension between $-0.5 t$ and $0.5 t$, 
where time $t$ starts when the two nuclei are fully overlapped in the 
longitudinal direction. 

To model the above effect in high energy density regions, we extend the
AMPT model to include the string melting mechanism
\cite{Lin:2001zk,Lin:2002gc,Lin:2003iq}, 
i.e, all excited strings, that are not projectile and target
nucleons without any interactions, are converted to partons
according to the flavor and spin structures of their valence quarks. 
In particular, a meson is converted to a quark and an antiquark, while 
a baryon is first converted to a quark and a diquark with weights
according to relations from the SU(6) quark model \cite{Pavkovic:rf},
and the diquark is then decomposed into two quarks. The quark and diquark
masses are taken to be the same as in the PYTHIA program 
\cite{Sjostrand:1993yb}, e.g. $m_u=5.6$ MeV$/c^2$, $m_d=9.9$
MeV$/c^2$, and $m_s=199$ MeV$/c^2$. We further assume that the above 
two-body decomposition is isotropic in the rest frame of the parent 
hadron or diquark, and the resulting partons do not undergo scatterings
until after a formation time given by $t_f=E_H/m_{T,H}^2$, with $E_H$
and $m_{T,H}$ denoting the energy and transverse mass of the parent hadron.  
Similar to minijet partons in the default AMPT model, initial positions of 
the partons from melted strings are calculated from those of their
parent hadrons using straight-line trajectories. 

The above formation time for partons is introduced to represent 
the time needed for their production from strong color fields. 
Although we consider hadrons before string melting as a convenient step in
modeling the string melting process, choosing a formation time that
depends on the momentum of the parent hadron ensures that partons from 
the melting of same hadron would have the same formation time. The advantage of
this choice is that the AMPT model with string melting reduces 
to HIJING results in the absence of partonic and hadronic interactions 
as these partons would then find each other as closest partners at 
the same freeze-out time and thus coalesce back to the original hadron. 
We note that the typical string fragmentation time of about 
1 fm/$c$ is not applied to the melting of strings as the 
fragmentation process involved here is considered just as an
intermediate step in modeling parton production from the energy field
of the strings in an environment of high energy density. 

\subsection{Parton cascade}\label{zpc}

In the transport approach, interactions among partons are
described by equations of motion for their Wigner distribution functions, 
which describe semi-classically their density distributions in phase space.
These equations can be approximately written as 
the following Boltzmann equations: 
\ber
&&p^\mu\partial_\mu f_a({\bf x},{\bf p},t)\nonumber\\
&&=\sum_m\sum_{b_1,b_2,\cdots,b_m}
\int\prod_{i=1}^m\frac{d^3p_{b_i}}{(2\pi)^32E_{b_i}}
f_{b_i}({\bf x},{\bf {p_{b_i}}},t) \nonumber \\
&&\times\sum_n\sum_{c_1,c_2,\cdots,c_n}
\int\prod_{j=1}^n\frac{d^3p_{c_j}}{(2\pi)^32E_{c_j}}|M_{m\rightarrow n}|^2
\nonumber \\
&&\times (2\pi)^4\delta^4\left(\sum_{k=1}^mp_{b_k}-\sum_{l=1}^np_{c_l}\right)
\nonumber \\
&&\times\left[-\sum_{q=1}^m\delta_{ab_q}\delta^3({\bf p}-{\bf {p_{b_q}}})
+\sum_{r=1}^n\delta_{ac_r}\delta^3({\bf p}-{\bf {p_{c_r}}})\right].  
\eer
In the above, $f_a({\bf x},{\bf p},t)$ is the distribution function of
parton type $a$ at time $t$ in the phase space, and $M_{m\rightarrow n}$ 
denotes the matrix element of the multi-parton interaction $m\rightarrow n$. 
If one considers only two-body interactions, these equations reduce to  
\ber
p^\mu\partial_\mu f({\bf x},{\bf p},t) \propto 
\int \sigma f({\bf {x_1}},{\bf {p_1}},t) f({\bf {x_2}},{\bf {p_2}},t),
\label{twobody}
\eer
where $\sigma$ is the cross section for partonic two-body scattering,
and the integral is evaluated over the momenta of other three partons 
with the integrand containing factors such as a delta function 
for momentum conservation. 

The Boltzmann equations are solved using Zhang's parton cascade (ZPC) 
\cite{Zhang:1997ej}, in which two partons undergo scattering whenever
they approach each other with a closest distance smaller than $\sqrt
{\sigma/\pi}$.  At present, ZPC includes only parton two-body
scattering such as $gg \rightarrow gg$ with cross sections calculated
from the pQCD.  For gluon elastic scattering, the leading-order QCD gives
\ber
\frac {d\sigma_{gg}}{dt}&=&\frac{9\pi \alpha_s^2}{2s^2} 
\left (3-\frac {ut}{s^2}-\frac {us}{t^2}-\frac {st}{u^2} \right ) \nonumber \\
&\simeq& \frac{9\pi \alpha_s^2}{2} \left (\frac {1}{t^2}+\frac {1}{u^2}
\right ),
\eer
where $\alpha_s$ is the strong coupling constant, and $s$, $t$ and $u$ 
are standard Mandelstam variables for elastic scattering of two partons. 
The second line in the above equation is obtained by keeping only the
leading divergent terms. Since the scattering angle ranges from
0 to $\pi/2$ for identical particles, one then has 
\cite{Zhang:1997ej}
\ber
\frac {d\sigma_{gg}}{dt} \simeq \frac{9\pi \alpha_s^2}{2t^2},
\eer
if the scattering angle is between 0 and $\pi$. 

The singularity in the total cross section can be regulated by a 
Debye screening mass $\mu$, leading to 
\ber
\frac {d\sigma_{gg}}{dt}&\simeq&\frac{9\pi \alpha_s^2}{2(t-\mu^2)^2},
\nonumber \\
\sigma_{gg}&=&\frac{9\pi \alpha_s^2}{2\mu^2} \frac {1}{1+\mu^2/s}.
\eer
The screening mass $\mu$ is generated by medium effects and is thus
related to the parton phase-space density. For the partonic system expected 
to be formed in Au+Au collisions at RHIC, the value of $\mu$ is on the 
order of one inverse fermi \cite{Zhang:1997ej}.  For massless partons
in a plasma at temperature $T$, their average colliding energy
is $\sqrt s \sim \sqrt{18} T$,  
thus $\mu<\sqrt s$ for $\mu=3$ fm$^{-1}$ leads to the requirement 
$T>141$ MeV. Since $s>\mu^2$ generally holds in hot QGP, 
the following simplified relation between the total parton elastic
scattering cross section and the screening mass is used in the 
ZPC \cite{smu2} 
\ber
\sigma_{gg}&\approx&\frac{9\pi \alpha_s^2}{2\mu^2}.
\label{sgg}
\eer
A value of $3$ fm$^{-1}$ for the screening mass $\mu$ thus leads to 
a total cross section of about $3$ mb for the elastic scattering
between two gluons. By changing the value of the screening mass $\mu$,
different cross sections can be obtained, and this will be used in
studying the effect of parton cross sections in heavy ion collisions
at RHIC. This cross section is used in AMPT not only in the 
default model, which includes only scatterings of minijet gluons, but also 
in the string melting model, which only includes scatterings of 
quarks/antiquarks of all flavors. We have therefore neglected in the latter
case the difference between the Casimir factors for quarks and gluons.

We note that minijet partons produced from hard scatterings in the HIJING 
model can lose energy by gluon splitting and transfer their energies to 
nearby soft strings. In the AMPT model, this so-called jet quenching
in the HIJING model is replaced by parton scatterings in ZPC.
Since only two-body scatterings are included in ZPC, higher-order
contributions to the jet energy loss are still missing in the AMPT model.

\subsection{Hadronization}\label{hadron}

Two different hadronization mechanisms are used in the AMPT model for
the two different initial conditions introduced in Sec.~\ref{hijing}.
In the default AMPT model, minijets coexist with the remaining part of 
their parent nucleons, and they together form new excited strings after 
partonic interactions. Hadronization of these strings are described by 
the Lund string model. In the AMPT model with string melting, these 
strings are converted to soft partons, and their hadronization is based
on a simple quark coalescence model, similar to that in the ALCOR model 
\cite{alcorpapers}. 

\subsubsection{Lund string fragmentation for the default AMPT model}

Hadron production from the minijet partons and soft strings in the default
AMPT model is modeled as follows. After minijet partons stop interacting, 
i.e., after they no longer scatter with other partons, they are
combined with their parent strings to form excited strings, 
which are then converted to hadrons according to the Lund string 
fragmentation model \cite{Andersson:1983jt,Andersson:ia}.  In the Lund model 
as implemented in the JETSET/PYTHIA routine \cite{Sjostrand:1993yb}, 
one assumes that a string fragments into quark-antiquark pairs with a 
Gaussian distribution in transverse momentum. A suppression factor of
0.30 is further introduced for the production of strange
quark-antiquark pairs relative to that of light quark-antiquark
pairs. Hadrons are formed from these quarks and antiquarks by
using a symmetric fragmentation function 
\cite{Andersson:1983jt,Andersson:ia}. Specifically, 
the transverse momentum of a hadron is given by those of its constituent 
quarks, while its longitudinal momentum is determined by 
the Lund symmetric fragmentation function \cite{ablund}
\ber
f(z) \propto z^{-1} (1-z)^a \exp (-b~m_{\perp}^2/z), 
\label{lundsf}
\eer
with $z$ denoting the light-cone momentum fraction of the produced hadron
with respect to that of the fragmenting string.  The average squared 
transverse momentum is then given by 
\ber
\langle \pp^2 \rangle
&=&\frac{\int \pp^2 f(z) d^2 \pp dz} {\int f(z) d^2 \pp dz} \nonumber \\
&=&\frac{\int_0^{z_{max}} z (1-z)^a \exp (-b~m^2/z) dz}
{b \int_0^{z_{max}} (1-z)^a \exp (-b~m^2/z) dz}.
\eer
For massless particles, it reduces to  
\ber
\langle \pp^2 \rangle=\frac{1}{b} \frac{\int_0^1 z (1-z)^a dz}
{\int_0^1 (1-z)^a dz}=\frac{1}{b(2+a)}.
\label{lundpt2}
\eer

Since quark-antiquark pair production from string fragmentation in the
Lund model is based on the Schwinger mechanism \cite{Schwinger:tp} for
particle production in strong field, its production probability is 
proportional to $\exp (-\pi m_\perp^2/\kappa)$, where $\kappa$ is the 
string tension, i.e., the energy in a unit length of string. Due to its 
large mass, strange quark production is suppressed by the factor
$e^{-\pi (m_s^2-m_u^2)/\kappa}$, compared to that of light quarks. 
Also, the average squared transverse momentum of produced particles is
proportional to the string tension, i.e., 
$\langle \pp^2 \rangle \propto \kappa$. Comparing this with 
Eq.~(\ref{lundpt2}), one finds that the two parameters $a$ and $b$ in 
the Lund fragmentation function are approximately related to the string 
tension by
\ber\label{tension}
\kappa\propto \frac {1}{b(2+a)}.
\eer

After production from string fragmentation, hadrons are given an 
additional proper formation time of 0.7 fm$/c$ \cite{ftime}. 
Positions of formed hadrons are then calculated from those of their parent 
strings by following straight-line trajectories.

\subsubsection{Quark coalescence for the AMPT model with string melting}
\label{pt-sm}

After partons in the string melting scenario stop interacting, we
model their hadronization via a simple quark coalescence model by
combining two nearest partons into a meson and three nearest quarks
(antiquarks) into a baryon (antibaryon). Since the invariant mass
of combined partons forms a continuous spectrum instead of a discrete
one, it is generally impossible to conserve 4-momentum when partons 
are coalesced into a hadron.  At present, we choose to conserve
the three-momentum during coalescence and determine the hadron species 
according to the flavor and invariant mass of coalescing partons
\cite{energy}. For pseudo-scalar and vector mesons with same flavor 
composition, the meson with mass closer to the invariant mass of 
coalescing quark and antiquark pair is formed. E.g., whether a $\pi^-$ 
or a $\rho^-$ is formed from the coalescence of a pair of $\bar u$ 
and $d$ quarks depends on if the invariance mass of the quarks is 
closer to the $\pi^-$ mass or the centroid of $\rho$ mass. The same
criterion applies to the formation of octet and decuplet baryons
that have same flavor composition. It is more complicated to treat the 
formation probabilities of flavor-diagonal mesons such as $\pi^0$ 
and $\eta$ in the pseudo-scalar meson octet, and $\rho^0$ and $\omega$ in the
vector meson octet. Neglecting the mixing of $\eta$ meson with the 
$s\bar s$ state, we take the following approach for these flavor-diagonal
mesons within the SU(2) flavor space. For $\pi^0$ formation from a 
$u\bar u$ or $d\bar d$ pair, the probability $P_{\pi^0}$ is determined
from the average of the numbers of formed $\pi^+$ and $\pi^-$ mesons 
divided by the total number of $u\bar u$ and $d\bar d$ pairs. 
Thus the total number of $\pi^0$, $n_{\pi^0}$, is determined
by applying the probability $P_{\pi^0}$ to each $u\bar u$ or $d\bar d$ pair.
The probability $P_{\rho^0}$ for forming a $\rho^0$ meson from a
$u\bar u$ or $d\bar d$ pair is determined by a similar procedure.
After sorting all $u\bar u$ or $d\bar d$ pairs according to their
invariant masses, the lightest $n_{\pi^0}$ pairs are assigned as
$\pi^0$ mesons. The rest of $u\bar u$ or $d\bar d$ pairs form $\rho^0$ mesons 
according to the probability of $P_{\rho^0}/(1-P_{\pi^0})$, and 
the remaining pairs form $\omega$ and $\eta$ mesons with equal probabilities.

The above quark coalescence model includes the formation of all mesons
and baryons listed in the HIJING program \cite{Gyulassy:ew}
except $\eta^\prime$, $\Sigma^*$ and $\Xi^*$, which are not present in
our hadronic transport model as well as $K_S^0$ and $K_L^0$ states.
The resulting hadrons are given an additional formation time of 
$0.7$ fm/$c$ in their rest frame before they are allowed to scatter 
with other hadrons during the hadron cascade. As partons freeze out 
dynamically at different times in the parton cascade, hadron formation 
from their coalescence thus occurs at different times, leading to 
the appearance of a coexisting phase of partons and hadrons during 
hadronization. 

\subsection{Hadron cascade}\label{art}

In the AMPT model, the following hadrons with all possible charges
are explicitly included: $\pi$, $\rho$, $\omega$, $\eta$, $K$, $K^*$,
and $\phi$ for mesons; $N$, $\Delta$, $N^*(1440)$, $N^*(1535)$, $\Lambda$,
$\Sigma$, $\Xi$, and $\Omega$ for baryons and corresponding antibaryons. 
Many other higher resonances are taken into account implicitly as 
intermediate states in scatterings between the above particles 
\cite{Li:1995pr,Li:2001xh}. Interactions among these hadrons and 
corresponding inverse reactions are included as discussed in 
following subsections.

\subsubsection{The ART model}

Hadron cascade in the AMPT model is based on the ART model 
\cite{Li:1995pr,Li:xh}, which is a relativistic transport model originally 
developed for heavy ion collisions at AGS energies. The ART model 
includes baryon-baryon, baryon-meson, and meson-meson elastic and
inelastic scatterings. It treats explicitly the isospin degrees of
freedom for most particle species and their interactions, making it
suitable for studying isospin effects in heavy ion collisions 
\cite{Li:1997px}. Since it includes mean-field potentials for 
nucleons and kaons, the ART model can also be used for studying 
the effect due to the hadronic equation of state. Resonances such 
as $\rho$ and $\Delta$ are formed from pion-pion and pion-nucleon 
scattering, respectively, with cross sections given by the standard 
Breit-Wigner form, and they also decay according to their respective 
widths. In all calculations presented in this study, the masses and 
widths of resonances are taken to be their values in the vacuum, 
i.e., effects due to possible modifications in dense hadronic matter 
\cite{rhopapers} 
are neglected. Also, we have turned off the potentials in the AMPT 
model as their effects are much less important than scatterings 
in high energy heavy ion collisions such as at SPS and RHIC.

For baryon-baryon scatterings, the ART model includes the following 
inelastic channels: $NN \lra N (\Delta N^*)$, 
$NN \lra \Delta (\Delta N^*(1440))$, $NN \lra NN (\pi \rho \omega)$, 
$(N \Delta) \Delta \lra N N^*$, and $\Delta N^*(1440) \lra N N^*(1535)$. 
In the above, $N^*$ denotes either $N^*(1440)$ or $N^*(1535)$,  
and the symbol $(\Delta N^*)$ denotes a $\Delta$ or an $N^*$, 
Also included are reaction channels relevant for kaon production, i.e., 
$(N \Delta N^*) (N \Delta N^*) \rightarrow (N \Delta) (\Lambda \Sigma) K$. 
Details on their cross sections and the momentum dependence of 
resonance widths can be found in the original ART model \cite{Li:1995pr}.

For meson-baryon scatterings, the ART model includes the following 
reaction channels for the formation and decay of resonances: 
$\pi N \lra (\Delta N^*(1440)~ N^*(1535))$, and $\eta N \lra N^*(1535)$. 
There are also elastic scatterings such as 
$(\pi \rho) (N \Delta N^*) \rightarrow (\pi \rho) (N \Delta N^*)$. 
As an example, the cross section for the elastic scattering 
of $\pi^0 N$ is evaluated by including heavier baryon resonances with masses 
up to 2.0 GeV$/c^2$ as intermediate states using the Breit-Wigner form but 
neglecting interferences between the amplitudes from different
resonances \cite{Li:1995pr}. The ART model further includes inelastic
reaction channels such as $\pi N \lra (\pi \rho \eta) \Delta$ and kaon
production channels such as 
$(\pi \rho \omega \eta) (N \Delta N^*) \lra K (\Lambda \Sigma)$. 
Kaon elastic scatterings with nucleons and baryon resonances are included 
with a constant cross section of 10 mb \cite{Li:1995pr}.  Antikaon 
elastic scatterings with nucleons and inelastic channels, such as 
$\bar K (N \Delta N^*) \lra \pi (\Lambda \Sigma)$, are 
included \cite{Song:1998id} using parameterized experimental data 
\cite{Cugnon:xw}. Also included are kaon production channels involving 
three-body final states, 
$(\pi \rho \omega) (N \Delta N^*) \rightarrow K \bar K N$ \cite{Song:1998id}.
Because of the difficulty associated with the three-body kinematics, 
the inverse kaon annihilation reactions of the above channels are neglected. 

For meson-meson interactions, the ART model includes both elastic and 
inelastic $\pi\pi$ interactions, with the elastic cross section consisting 
of $\rho$ meson formation and the remaining part treated as elastic 
scattering. Kaon production from inelastic scatterings of light mesons is  
included via the the reactions $(\pi \eta)(\pi \eta)\lra K \bar K$ and 
$(\rho \omega)(\rho \omega)\lra K \bar K$. Kaon or antikaon elastic 
scatterings with mesons in the SU(2) multiplets except the pion are 
included using a constant cross section of 10 mb \cite{Li:1995pr}, 
while the kaon-pion elastic scattering is modeled through the $K^*$ 
resonance \cite{Lin:2001cx}. 

\subsubsection{Explicit inclusion of $K^*$ mesons}
\label{exkstar}

The original ART model \cite{Li:1995pr} includes the $K^*$ resonance 
implicitly through elastic $\pi K$ scattering with the standard 
Breit-Wigner form for the cross section \cite{Ko:1981th}, i.e., 
$\sigma_{\pi K}=60 {\rm ~mb}/\left [ 1+4 (\sqrt s-m_{K^*})^2/
\Gamma_{K^*}^2 \right ]$.
Since the $K^*$ meson not only enhances elastic scattering between pion
and kaon but also adds to strange particle production through its
addition to the strangeness degeneracy, which becomes important when
the hadronic matter is highly excited, $K^*$ and $\bar {K^*}$ are
included explicitly in the hadronic phase of the AMPT model 
\cite{Lin:2001cx}. In addition to its formation from $\pi K$
scattering and its decay, elastic scatterings of $K^*$ with 
$(\rho \omega \eta)$ are included using same constant cross 
section of 10 mb as those for kaons. Inelastic reaction channels of 
$(\pi \eta) (\rho \omega) \lra K^* \bar K$ or $\bar {K^*} K$ and  
$\pi K \lra K^* (\rho \omega)$ are also included \cite{Brown:1991ig}.

\subsubsection{Baryon-antibaryon annihilation and production}
\label{bbmm}

In heavy ion collisions at or above SPS energies, antibaryon production 
becomes significant and needs to be treated explicitly during hadron 
cascade. The AMPT model initially only includes $N \bar N$ annihilation 
\cite{Zhang:2000bd}. It was later extended to include 
$(N\Delta N^*) (\bar N\bar\Delta\bar N^*)$ annihilation and also the inverse 
reactions of baryon-antibaryon pair production from mesons \cite{Lin:2001cx}.

The total cross section for $p \bar p$ annihilation is known empirically,
and the data has been parameterized as \cite{gjwang} 
\ber
\sigma_{p\bar p}=67{\rm ~mb}/p_{\rm lab}^{0.7},
\eer
where $p_{\rm lab}$ in GeV$/c$ is the proton momentum in the rest frame
of the antiproton. Following Ref.~\cite{Li:1995pr}, a maximum cross 
section of 400 mb is imposed at low $p_{\rm lab}$. Using 
phase-space considerations \cite{Ko:hf}, the branching ratios of 
$p\bar p$ annihilation to different multi-pion states are determined
according to \cite{Ko:hf} 
\ber
M_n(\sqrt s)=C \left [\frac {1}{6\pi^2} 
\left ( \frac {\sqrt s}{m_\pi} \right )^3 \right ]^n 
\frac {(4n-4)! (2n-1)}{(2n-1)!^2 (3n-4)!},
\eer
where $\sqrt s$ is the center-of-mass energy of the proton and antiproton, 
and $n$ is the number of pions in the final state.  The dominant final 
states at moderate energies then involve several pions.  For example, the
branching ratios at $p_{\rm lab}=4$ GeV$/c$ are 0.033, 0.161, 0.306, 
0.286, 0.151, 0.049 and 0.011, respectively, for $n$ from 3 to 9. 
For baryon-antibaryon annihilation channels involving baryon 
resonances $\Delta$ or $N^*$, their annihilation cross sections 
and branching ratios are taken to be the same as for $p \bar p$ 
annihilation at same center-of-mass energy.  

To include the inverse reactions that produce baryon-antibaryon pairs 
during hadron cascade, which currently only treats scattering of two 
particles, we have further assumed that the final state of three pions 
is equivalent to a $\pi \rho$ state; the four-pion final state is
equivalent to $\rho \rho$ and $\pi \omega$ with equal probabilities; 
the five-pion state is equivalent to $\rho \omega$; and the six-pion
state is equivalent to $\omega \omega$. The cross sections for 
baryon-antibaryon pair production from two mesons are then obtained
from detailed balance relations. As shown later in the paper 
(Fig.~\ref{na49-dndy-nofsi}), the above approximate treatment of 
antibaryon annihilation and production via two-meson states gives
a satisfactory description of measured antiproton yield in central Pb+Pb 
collisions at SPS.

\subsubsection{Multistrange baryon production from strangeness-exchange 
reactions}
\label{ms}

\begin{figure}[ht]
\centerline{\epsfig{file=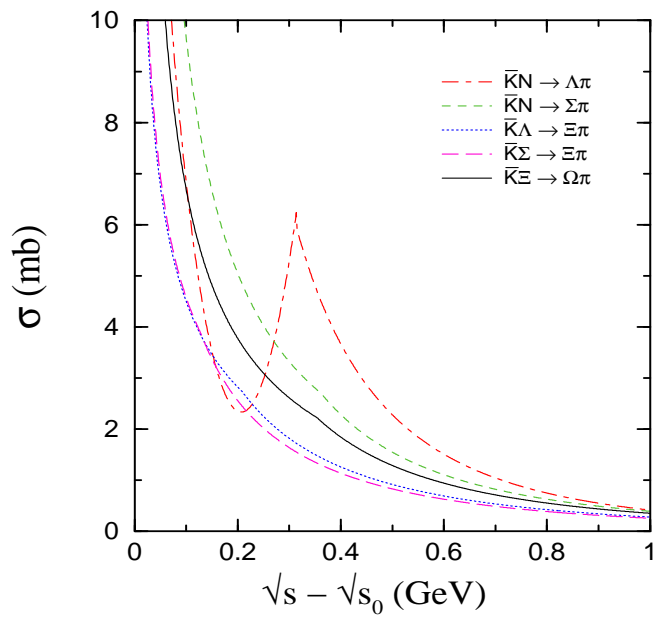,width=3.4in,height=3.2in,angle=0}}
\caption{(Color online) Isospin-averaged cross sections for (multi)strange 
baryon production as functions of center-of-mass energy of interacting
antikaon and nucleon/hyperon.}
\label{cross}
\end{figure}

Productions of multistrange baryons such as $\Xi$ and $\Omega$  
are included in the AMPT model  \cite{Pal:2001zw} through 
the strangeness-exchange reactions such as  
\ber
\bar K  (\Lambda \Sigma) \lra \pi \Xi, ~ \bar K  \Xi \lra \pi \Omega, 
\eer
Since there is no experimental information on their cross sections, we 
have assumed that the matrix elements for 
$\bar K  (\Lambda \Sigma) \rightarrow \pi \Xi$ and 
$\bar K  \Xi \rightarrow \pi \Omega$ are the same as that for the 
reaction $\bar K N \rightarrow \pi \Sigma$ \cite{Koch:1986ud} at the same 
amount of energy above corresponding thresholds. The isospin-averaged
cross section for the reaction $\bar K N \rightarrow \pi \Sigma$ 
can be related to the cross sections for the reactions $K^-p\to\Sigma^0\pi^0$ 
and $K^-n\to\Sigma^0\pi^-$, which are known empirically and have
been parameterized in Ref.~\cite{Cugnon:xw}, by
\ber
\sigma_{\bar KN\to\Sigma\pi}&=&\frac{3}{2}(\sigma_{K^-p\to\Sigma^0\pi^0}
+\sigma_{K^-n\to\Sigma^0\pi^-}).  
\eer
In Fig.~\ref{cross}, this cross section as well as the
isospin-averaged cross sections for other multistrange baryon production
reactions are shown as functions of the center-of-mass energy above
the threshold values of the interacting antikaon and nucleon/hyperon.
The cross sections for the inverse multistrange baryon destruction 
reactions are then determined by detailed balance relations.  We note 
that these cross sections are comparable to those predicted by the
coupled-channel calculations based on the SU(3) invariant hadronic
Lagrangian with empirical masses and coupling constants
\cite{Li:2002yd}. With these cross sections, the ART
transport model is able to describe the measured $\Xi$ production in
heavy ion collisions at the AGS energies \cite{Pal:2004kh}. 
We note that, for strange baryons $\Lambda$, $\Xi$, $\Omega$ and their 
anti-particles, only their interactions with mesons have been included, while 
their annihilations by baryons have not been included in the AMPT model 
at present. 

\subsubsection{$\phi$ meson production and scattering}

The AMPT model also includes $\phi$ meson formation from and decay to 
kaon-antikaon pair with the formation cross section given by 
the standard Breit-Wigner form \cite{Pal:2002aw}.  Inelastic scatterings 
of the phi meson include baryon-baryon channels, 
$(N \Delta N^*) (N \Delta N^*) \rightarrow \phi NN$, and meson-baryon 
channels, $(\pi \rho) (N \Delta N^*) \lra \phi (N \Delta N^*)$,   
where the cross sections for the forward-going reactions are taken 
from the one-boson-exchange model \cite{Chung:1997mp}. The meson-baryon 
channels also include $K (\Lambda \Sigma) \lra \phi N$ with cross 
section taken from a kaon-exchange model \cite{Ko:kw}. 

Phi meson scatterings with mesons such as $\pi, \rho, K$ and $\phi$ 
have been studied before, and the total collisional width was
found to be less than 35 MeV$/c^2$ \cite{Ko:1993id}. A recent calculation 
based on the Hidden Local Symmetry Lagrangian \cite{Alvarez-Ruso:2002ib} 
shows, however, that the collisional rates of $\phi$ with pseudo-scalar 
($\pi$, $K$) and vector ($\rho$, $\omega$, $K^*$, $\phi$) mesons are 
appreciably larger. Assuming that the matrix elements are independent 
of center-of-mass energy, we have included all these possible 
reactions, i.e., $\phi (\pi \rho \omega) \lra (K K^*)(\bar K \bar {K^*})$, and 
$\phi (K K^*) \lra (\pi \rho \omega) (K K^*)$, with cross sections 
determined from the partial collisional widths given in 
Ref.~\cite{Alvarez-Ruso:2002ib}. The cross section for the elastic 
scattering of the $\phi$ meson with a nucleon is set to 8 mb 
while the $\phi$ meson elastic cross section with a meson is set to 5 mb.
The value of 8 mb is the $\phi N$ total cross section \cite{Ko:kw} 
estimated from the $\phi$ meson photoproduction data 
\cite{Behrend:1975xk} and thus represents the upper bound on the $\phi$ meson 
elastic cross section with a nucleon; quark counting then gives 5 mb 
as the upper bound on the $\phi$ meson elastic cross section with a meson.
Note that, in most calculations of our previous study on $\phi$ meson 
productions at SPS and RHIC \cite{Pal:2002aw}, 
the elastic cross section for $\phi$ meson scattering 
with a nucleon was taken to be 0.56 mb as extracted in Ref.~\cite{joos} 
using the vector meson dominance model and the older 
$\phi$ meson photoproduction data, while based on results of 
Ref.~\cite{Alvarez-Ruso:2002ib} the $\phi K$ elastic cross section 
was extracted to be about 2 mb \cite{Pal:2002aw},  
which was then used as the $\phi$ meson elastic cross section with 
other mesons \cite{phielastic}. 

\subsubsection{Other extensions}

Other extensions of the ART model have also been made in the hadronic 
phase of the AMPT model. Antibaryon resonances such as $\bar \Delta$ and 
$\bar N^*$ have been included explicitly with their formations, decays, 
and scatterings analogous to those of baryon resonances 
\cite{Pal:2001zw,Lin:2002gc}. Also, inelastic meson-meson collisions such 
as $\pi \pi \lra \rho \rho$ have been added, and elastic scatterings 
between $\pi$ and $(\rho \omega \eta)$ have been included with cross
sections taken to be 20 mb. To address chemical equilibration of
$\eta$ mesons, which affects the height (or the $\lambda$ parameter)
of the correlation functions in two-pion interferometry, inelastic
scatterings of $\eta$ meson with other mesons have also been included
with a constant cross section of 5 mb \cite{Lin:2002gc}, which is roughly 
in line with recent theoretical predictions based on the Hidden Local 
Symmetry Lagrangian \cite{Liu:2005jb}.

\section{Results for $pp$ and $p\bar p$ collisions}
\label{pp}

The default AMPT results with no popcorn mechanism (see
discussions on the popcorn mechanism in Sec.~\ref{popcorn}) for
$pp$ and $p\bar p$ collisions are essentially the same as the results
from the HIJING model. In this section, we compare the results from 
the default AMPT model with or without the popcorn mechanism against
available data from $pp$ and $p\bar p$ collisions \cite{amptpp},  
where HIJING values for the $a$ and $b$ parameters, $a=0.5$ and $b=0.9$ 
GeV$^{-2}$, are used in Eq.~(\ref{lundsf}) for the Lund fragmentation 
function.

\subsection{Rapidity distributions}

\begin{figure}[ht]
\centerline{\epsfig{file=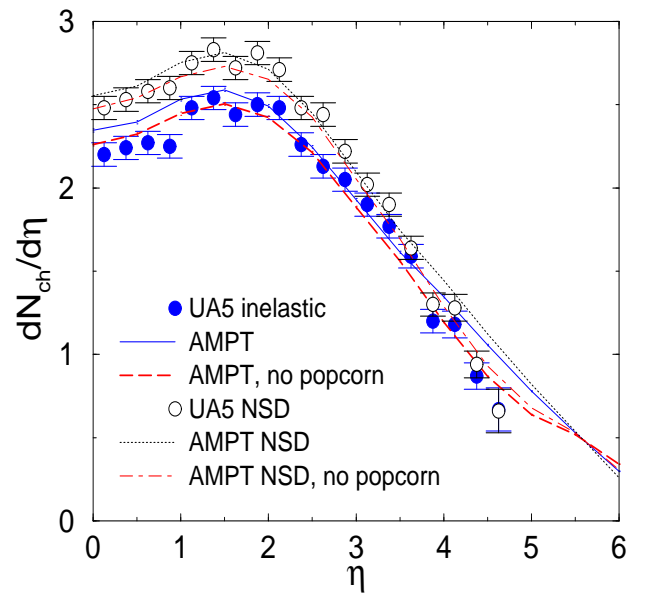,width=3.2in,height=3.2in,angle=0}}
\caption{(Color online) Pseudorapidity distributions of charged particles 
for $p\bar p$ collisions at $\sqrt s=200$ GeV.} 
\label{nch-ua5}
\end{figure}

\begin{figure}[htb]
\centerline{\epsfig{file=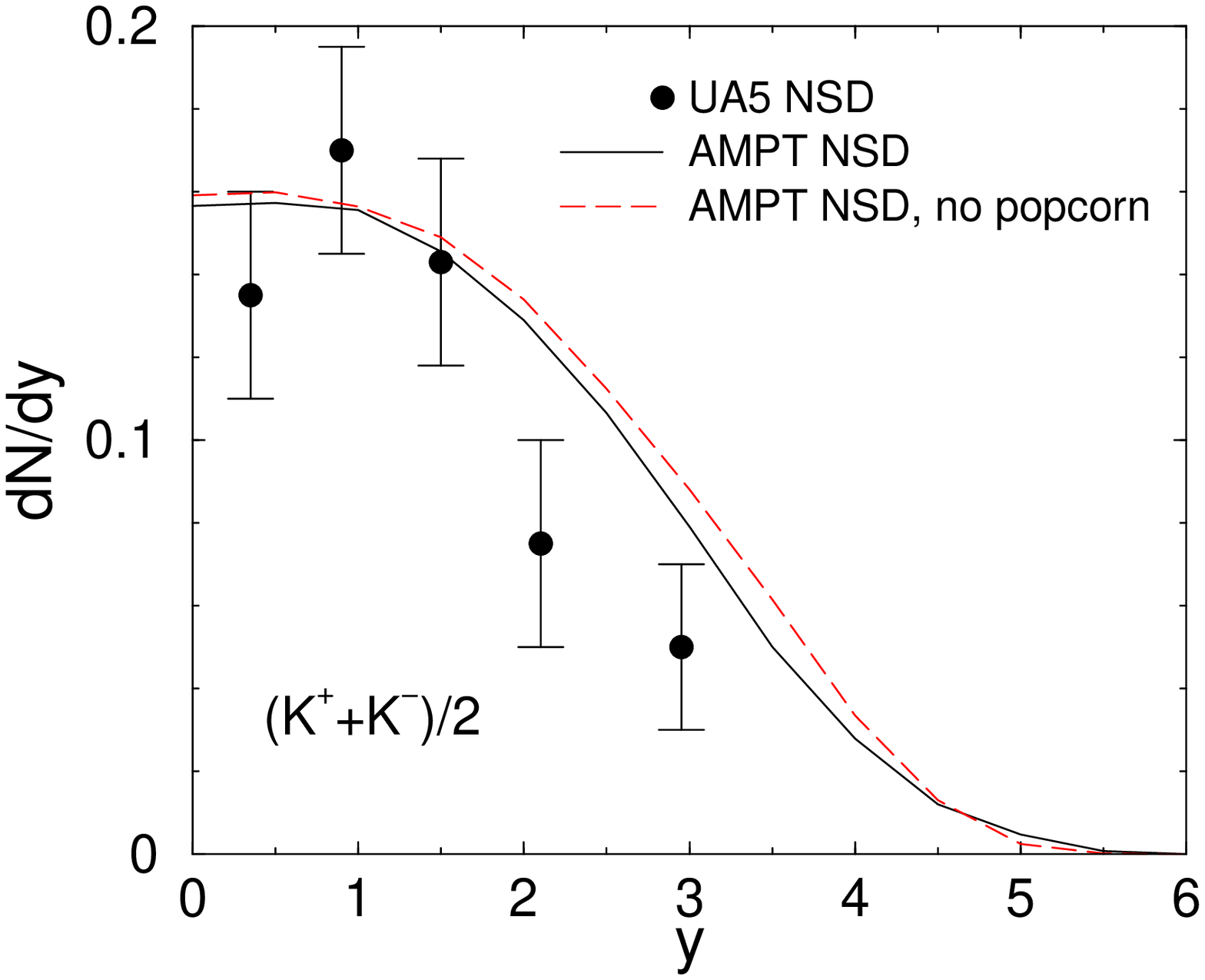,width=3.2in,height=3.2in,angle=0}}
\caption{(Color online) Same as Fig.~\ref{nch-ua5} for kaons.}
\label{kaon-ua5}
\end{figure}

In Fig.~\ref{nch-ua5}, the AMPT results on charged particle pseudorapidity 
distribution are compared with the UA5 data for $p\bar p$ collisions 
at $\sqrt s=200$ GeV \cite{Alner:1986xu}. Since the AMPT results given 
by the solid or the long-dashed curves represent all inelastic events 
with no trigger conditions, they should be compared with the UA5 
inelastic data.  The AMPT non-single-diffractive (NSD) results have 
included the UA5 NSD trigger by requiring events to have at least 
one charged particle each in both ends of the pseudorapidity 
intervals $2<|\eta|<5.6$ \cite{Alner:1986xu}. We see that the AMPT 
model with or without the popcorn mechanism agrees reasonably with both 
NSD and inelastic data. The kaon rapidity distribution from the AMPT 
model is compared with the UA5 NSD data in Fig.~\ref{kaon-ua5},
and both results with or without the popcorn mechanism also agree 
reasonably with the data. 

\begin{figure}[htb]
\centerline{\epsfig{file=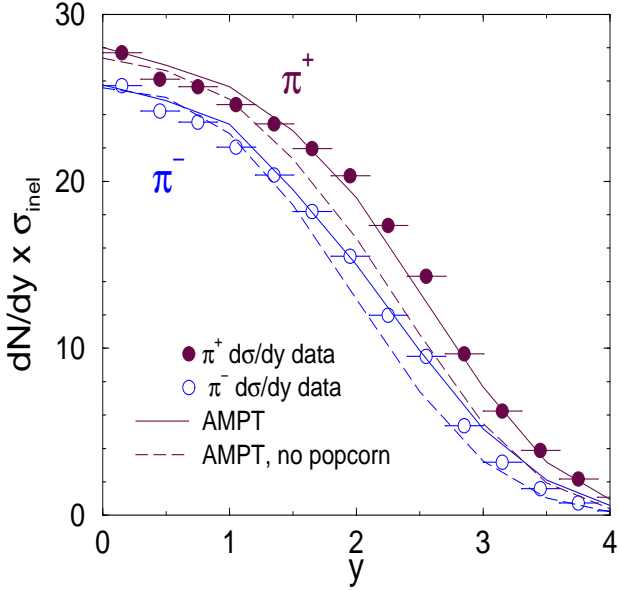,width=3.2in,height=3.2in,
angle=0}}
\caption{(Color online) Rapidity distributions of pions for 
$pp$ collisions at $P_{lab}=400$ GeV$/c$.
Circles are data from the LEBC-EHS
Collaboration \protect\cite{Aguilar-Benitez:1991yy}.}
\label{pion-pp400}
\end{figure}

\begin{figure}[htb]
\centerline{\epsfig{file=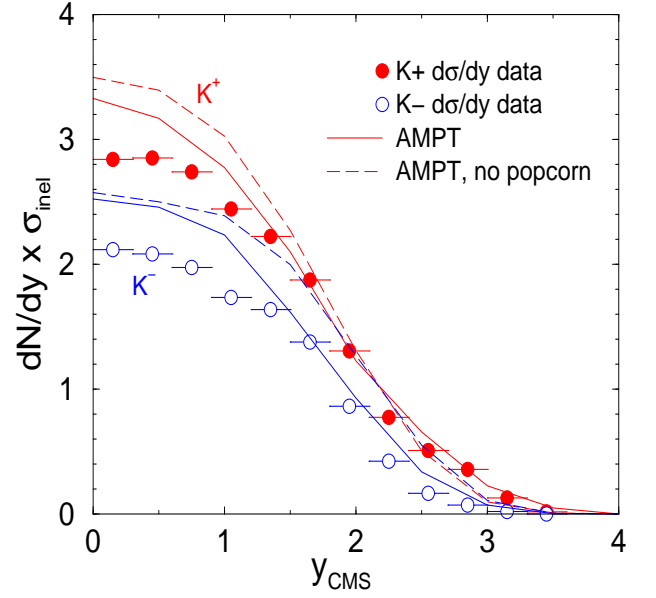,width=3.2in,height=3.2in,
angle=0}}
\caption{(Color online) Same as Fig.~\ref{pion-pp400} for kaons.}
\label{kaon-pp400}
\end{figure}

\begin{figure}[htb]
\centerline{\epsfig{file=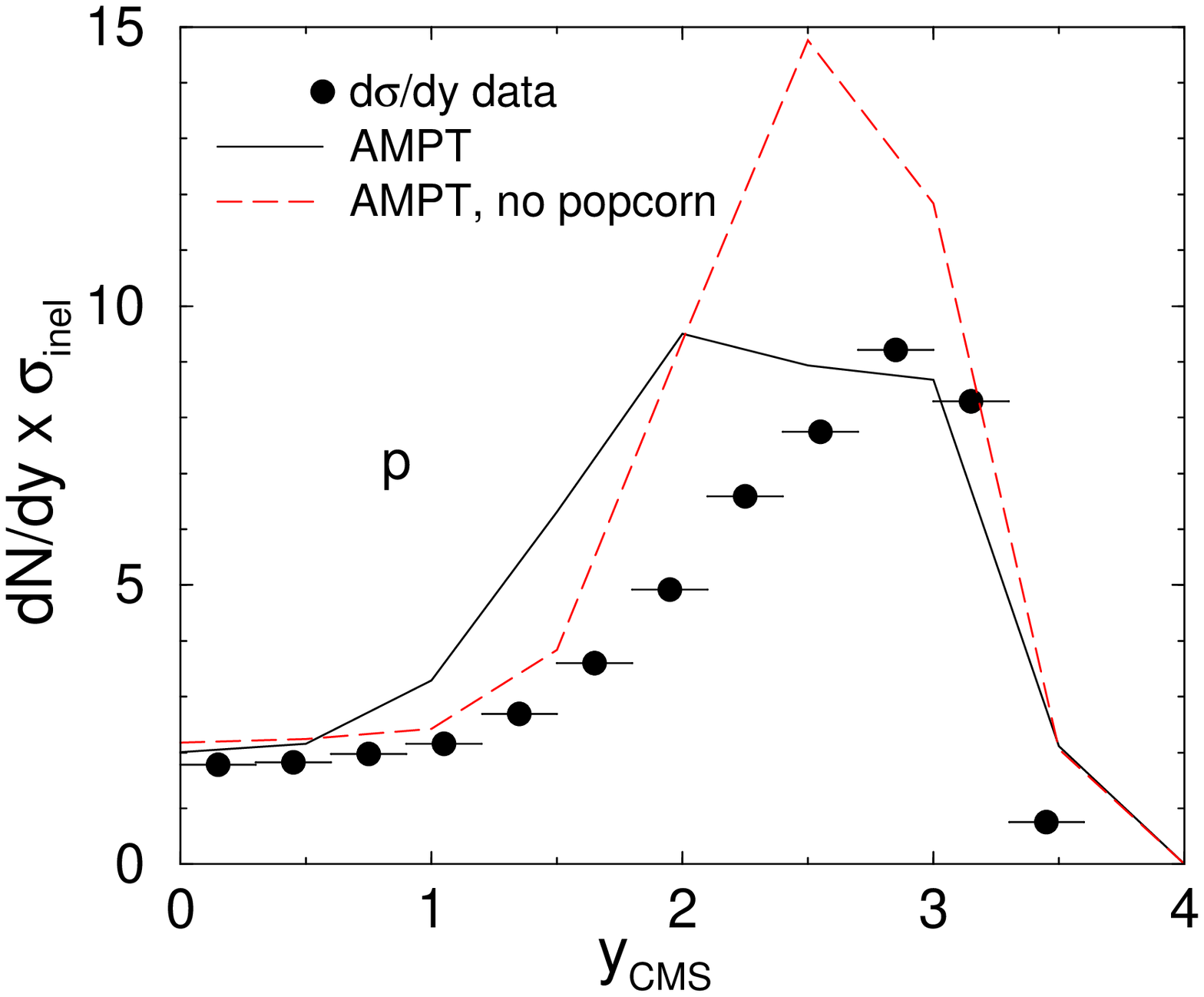,width=3.2in,
height=3.2in,angle=0}}
\caption{(Color online) Same as Fig.~\ref{pion-pp400} for protons.}
\label{pr-pp400}
\end{figure}

\begin{figure}[htb]
\centerline{\epsfig{file=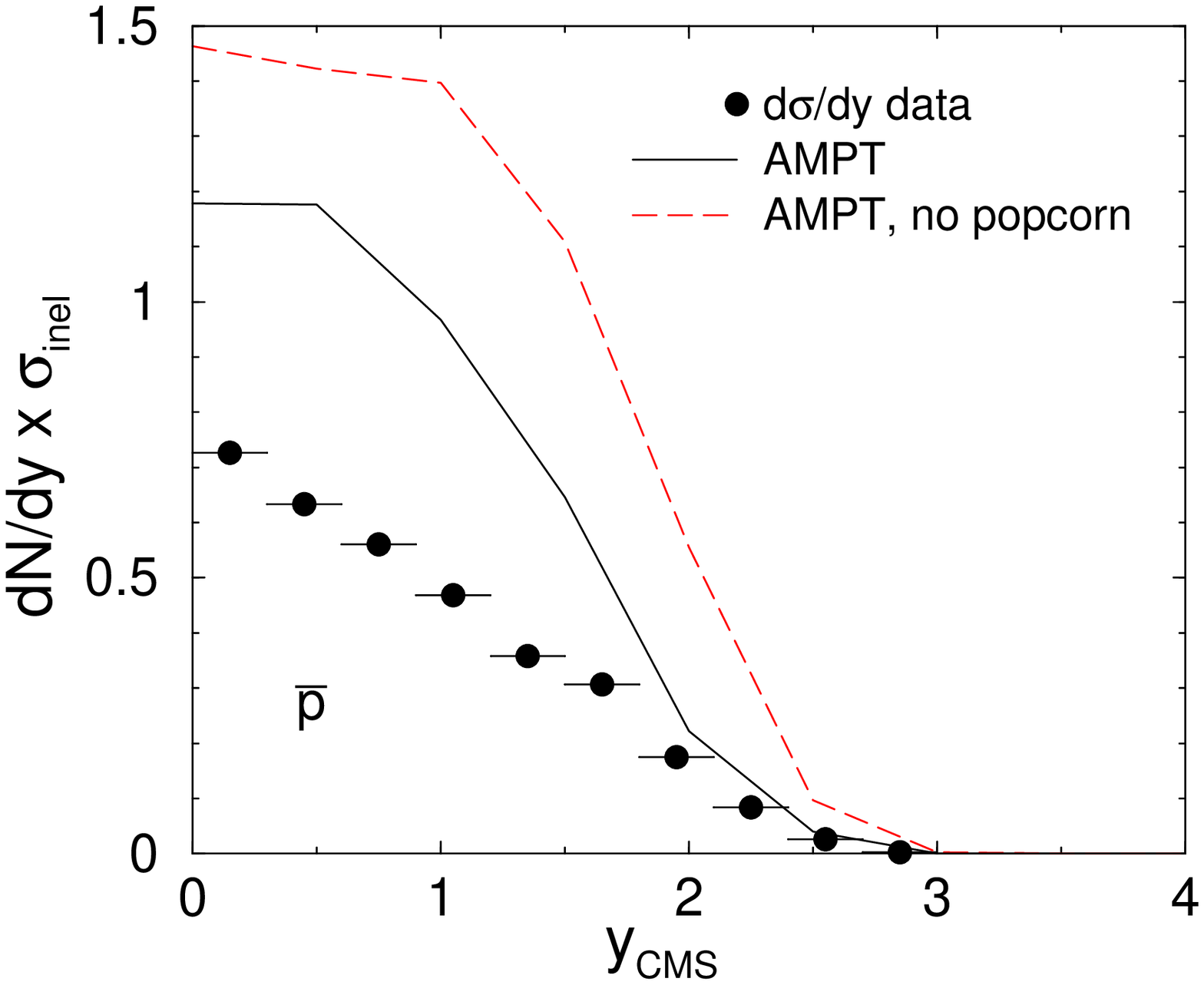,width=3.2in,height=3.2in,
angle=0}}
\caption{(Color online) Same as Fig.~\ref{pion-pp400} for antiprotons.}
\label{pbar-pp400}
\end{figure}

For $pp$ collisions at $P_{lab}=400$ GeV$/c$, the rapidity distributions 
of pions, kaons, protons and antiprotons are shown, respectively, 
in Fig.~\ref{pion-pp400}, Fig.~\ref{kaon-pp400}, Fig.~\ref{pr-pp400},
and Fig.~\ref{pbar-pp400}. Curves with circles are measured
cross sections from the LEBC-EHS Collaboration \cite{Aguilar-Benitez:1991yy}, 
while for comparison the AMPT results have been scaled up 
by the inelastic cross section at this energy ($\sigma_{inel}=32$ mb). 
We see that similar descriptions of the charged particle data 
are obtained with and without the popcorn mechanism. 
Including the popcorn mechanism gives, however, a better 
description of measured pion, kaon and antiproton yields and the shape of 
the proton rapidity distribution. We have thus included the popcorn 
mechanism in all AMPT calculations.

\subsection{Transverse momentum spectra}

\begin{figure}[htb]
\centerline{\epsfig{file=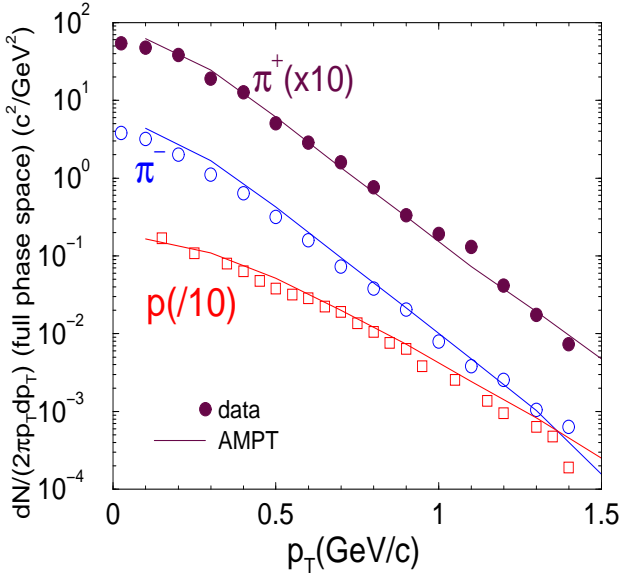,width=3.2in,height=3.2in,
angle=0}}
\caption{(Color online) Transverse momentum spectra of charged pions 
and protons for $pp$ collisions at $P_{lab}=24$ GeV$/c$ with data from 
Ref.~\protect\cite{Blobel:1973jc}.}
\label{pt-plab24}
\end{figure}

The transverse momentum spectra of charged pions and protons 
for $pp$ collisions at $P_{lab}=24$ GeV$/c$ from the AMPT model
are shown in Fig.~\ref{pt-plab24}. They are seen to reproduce reasonably 
well the experimental data from Ref.~\cite{Blobel:1973jc}.

\begin{figure}[htb]
\centerline{\epsfig{file=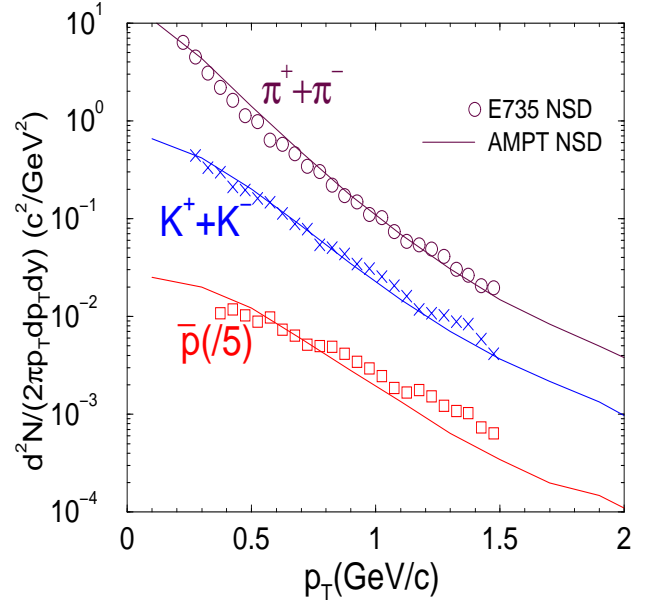,width=3.2in,height=3.2in,
angle=0}}
\caption{(Color online) Transverse momentum spectra 
of charged pions, charged kaons and 
antiprotons from $p\bar p$ collisions at $\sqrt s=1.8$ TeV.}
\label{pt-tevatron}
\end{figure}

At the Tevatron energy of $\sqrt s=1.8$ TeV, results from the AMPT model
for the transverse momentum spectra of pions, kaons and antiprotons 
are shown in Fig.~\ref{pt-tevatron} and compared with data from 
the E735 Collaboration~\cite{Alexopoulos:1990hn}.
The AMPT NSD results have included the trigger by 
requiring events to have at least one charged particle each in 
both ends of the pseudorapidity intervals $3<|\eta|<4.5$ 
\cite{Alexopoulos:1988na}. We see that measured momentum spectra 
except for antiprotons are reproduced reasonably well by the AMPT model. 
Note that the E735 $\pt$ spectrum data shown in 
Fig.~\ref{pt-tevatron} have been averaged over rapidity $y$ from
weighing each track by the rapidity range of the spectrometer
~\cite{Alexopoulos:1990hn},
thus they are for $d^2N/(2\pi \pt d\pt dy)$
instead for $d^2N/(2\pi \pt d\pt d\eta)$
even though the E735 spectrometer covers the acceptance of $-0.36<\eta<1.0$.

\subsection{Energy dependence}

\begin{figure}[htb]
\centerline{\epsfig{file=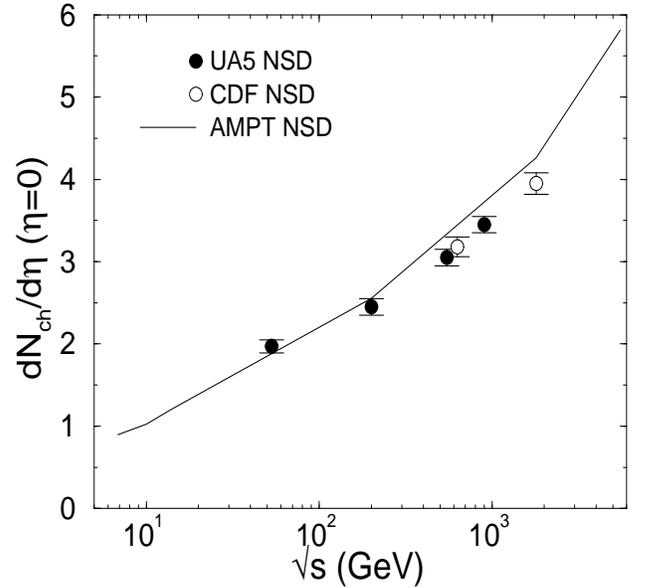,width=3.2in,height=3.2in,
angle=0}}
\caption{Energy dependence of $dN_{ch}/d\eta$ at mid-pseudorapidity 
for $pp$ and $p\bar p$ collisions.}
\label{dnchde-s}
\end{figure}

Fig.~\ref{dnchde-s} shows the energy dependence of 
$dN_{ch}/d\eta$ at $\eta=0$ for $pp$ and $p\bar p$ collisions 
together with UA5 and CDF data at the Tevatron \cite{Alner:1986xu,Abe:1989td}.
The AMPT NSD results above $\sqrt s=1$ TeV have included the CDF NSD trigger 
by requiring events to have at least one charged particle each in both ends 
of the pseudorapidity intervals of $3.2<|\eta|<5.9$ \cite{Abe:1989td},  
and the AMPT NSD results below $\sqrt s=1$ TeV have included the UA5 
NSD trigger by requiring events to have at least one charged particle 
each in both ends of the pseudorapidity intervals of $2<|\eta|<5.6$ 
\cite{Alner:1986xu}. The agreement with the Tevatron data is reasonable.

\begin{figure}[htb]
\centerline{\epsfig{file=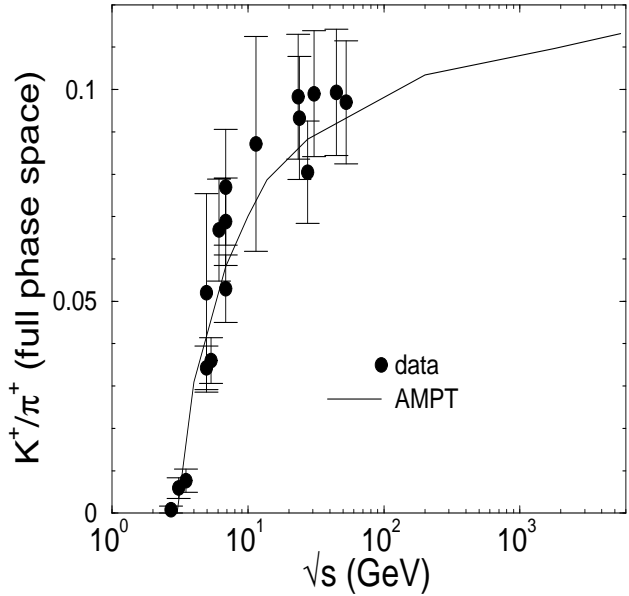,width=3.2in,height=3.2in,
angle=0}}
\caption{Energy dependence of the $K^+/\pi^+$ ratio in full 
phase space for $pp$ collisions.}
\label{kppip-s}
\end{figure}

The energy dependence of the full phase space $K^+/\pi^+$ ratio 
for $pp$ collisions is shown in Fig.~\ref{kppip-s} together with 
the data compiled in Fig.~7 of Ref.~\cite{Afanasiev:2002mx}. 
It is seen that the AMPT model reproduces the data reasonably well.

\begin{figure}[htb]
\centerline{\epsfig{file=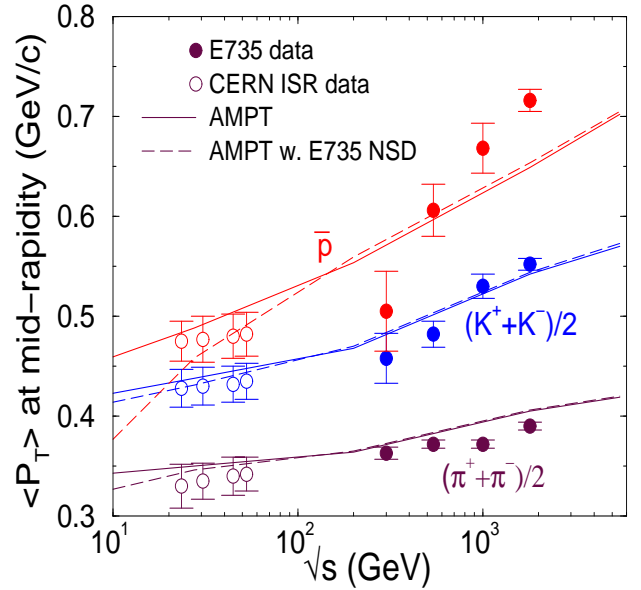,width=3.2in,height=3.2in,
angle=0}}
\caption{(Color online) Energy-dependence of the mean transverse momenta of 
pions, kaons and antiprotons for $pp$ and $p\bar p$ collisions.}
\label{meanpt-s}
\end{figure}

The energy dependence of the mean transverse momenta of pions, kaons 
and antiprotons are shown in Fig.~\ref{meanpt-s}, where filled circles 
represent the NSD data from the E735 Collaboration \cite{Alexopoulos:1993wt}  
and open circles represent data from the CERN Intersecting Storage Rings 
(ISR) \cite{Alper:1974rw}. Note that the AMPT NSD results shown in 
Fig.~\ref{meanpt-s} have included the E735 NSD trigger 
\cite{Alexopoulos:1988na} by requiring events to have at least one 
charged particle each in both ends of the pseudorapidity intervals 
$3<|\eta|<4.5$. However, the trigger condition for the CERN ISR data 
is different.  
From the comparison with the AMPT inelastic results (solid curves),
we see that NSD triggers have larger effects at lower energies.

\section{Results at SPS energies}
\label{sps}

To make predictions for heavy ion collisions at RHIC, we first use the
AMPT model to study heavy ion collisions at SPS. In particular, parameters
in the AMPT model are determined by fitting the experimental data from 
central Pb+Pb collisions at the laboratory energy of $158A$
GeV, corresponding to a center-of-mass energy of about 
$\sqrt {s_{NN}}=17.3$ GeV.

\subsection{Rapidity distributions}
\label{sps-mul}

\begin{figure}[ht]
\centerline{\epsfig{file=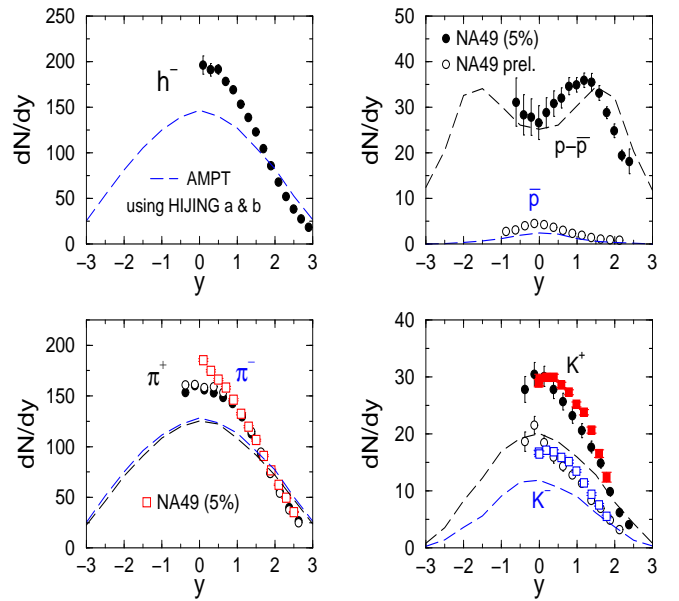,width=3.4in,height=3.2in,
angle=0}}
\caption{(Color online) 
Rapidity distributions in central ($b\leq 3$ fm) Pb+Pb collisions 
at $\sqrt {s_{NN}}=17.3$ GeV. Circles and squares are experimental data, 
while dashed curves are results from the AMPT model using
default $a$ and $b$ parameters as in the HIJING model.}
\label{na49-dndy-hjab}
\end{figure}

In Fig.~\ref{na49-dndy-hjab}, we show the rapidity distributions 
of negatively charged particles (upper left panel), net-protons and
antiprotons (upper right panel), charged pions (lower left panel), and
charged kaons (lower right panel) in central ($b\leq 3$ fm) Pb+Pb
collisions at SPS, obtained from the AMPT model with the default
values of $a=0.5$ and $b=0.9$ GeV$^{-2}$ in the HIJING model
for the string fragmentation function.  Compared with experimental 
data for 5\% most central Pb+Pb collisions from the NA49 Collaboration 
\cite{Appelshauser:1998yb,Bachler:hu,Afanasiev:2002mx}, 
it is seen that the AMPT model 
with these $a$ and $b$ parameters under-predicts both the negatively 
charged particle \cite{Zhang:1999mq} and kaon multiplicities.

\begin{figure}[ht]
\centerline{\epsfig{file=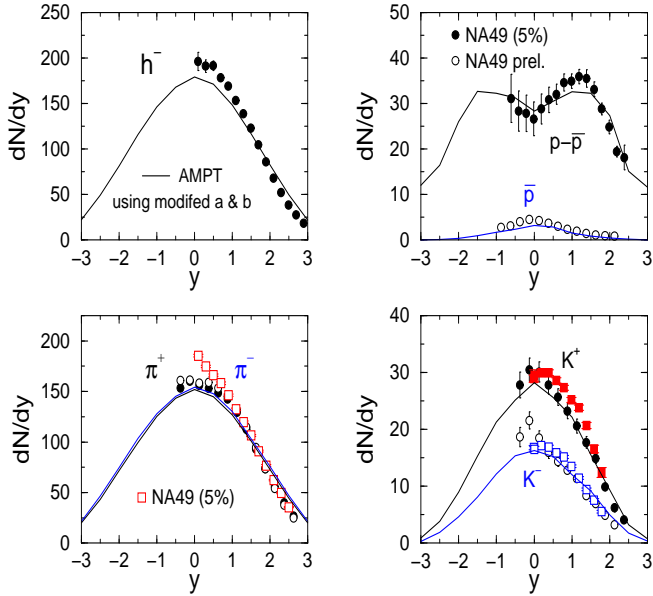,width=3.4in,height=3.2in,
angle=0}}
\caption{(Color online) 
Same as Fig.~\ref{na49-dndy-hjab} but with solid curves for
results from the AMPT model using modified $a$ and $b$ parameters in
the Lund string fragmentation function.}
\label{na49-dndy-newab}
\end{figure}

To increase the particle multiplicity in the AMPT model, we vary the
$a$ and $b$ parameters, and find that the experimental data can be
reasonably reproduced with $a=2.2$ and $b=0.5$ GeV$^{-2}$ as shown 
in Fig.~\ref{na49-dndy-newab}.  As seen from Eq.~(\ref{lundsf}), 
a larger value of $a$ corresponds to a softer fragmentation function, i.e.,
it leads to a smaller average transverse momentum for produced
hadrons and thus increases the particle multiplicity. The modified
parameters also affect the string tension. According to Eq.~(\ref{tension}), 
it is now 7\% larger, and this leads to a strangeness suppression
factor of 0.33 instead of the default value of 0.30.  Since the HIJING
model with default $a$ and $b$ parameters reproduces the charged particle
multiplicities in $pp$ and $p\bar p$ collisions, the AMPT model with these
parameters can probably describe peripheral heavy ion collisions. 
The $a$ and $b$ values in the AMPT model is thus expected to depend on 
the centrality of heavy ion collisions, but this has not been studied.

\begin{figure}[ht]
\centerline{\epsfig{file=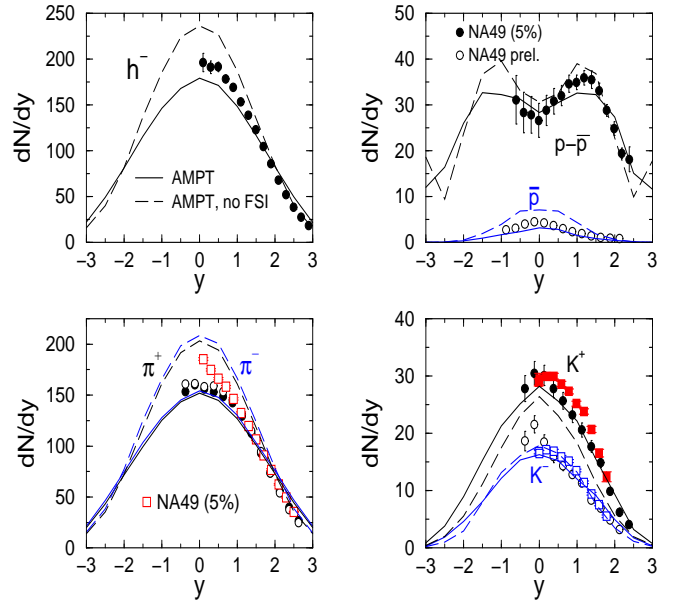,width=3.4in,
height=3.2in,angle=0}}
\caption{(Color online) 
Same as Fig.~\ref{na49-dndy-newab} but with dashed curves for
results from the AMPT model without partonic and hadronic final-state
interactions.}
\label{na49-dndy-nofsi}
\end{figure}

Since the probability for minijet production in the HIJING model 
is very small in collisions at SPS energies (only about 4 minijet partons 
for a central Pb+Pb event on average), the partonic stage in the default 
AMPT model does not play any role for most observables in these collisions. 
Final-state hadronic scatterings are, however, important, and their 
effects are illustrated in Fig.~\ref{na49-dndy-nofsi}. It is seen that
final-state interactions reduce the pion and antiproton yields, but
increase strangeness production, e.g., the sum of $K^-$ and $\Lambda$
as well as the sum of $K^+$ and $\bar\Lambda$ are both increased by about
20\%. Overall, the kaon yield in the default AMPT model is larger than
that in the HIJING model by about 40\% due to combined effects of 
modified Lund string fragmentation and final-state interactions.
 
\subsection{Baryon stopping and antiproton production}
\label{popcorn}

Understanding baryon stopping in relativistic heavy ion collisions 
is important as it is related to the total energy deposited in the
produced hot dense matter during the collisions. The observed
relatively large baryon stopping in these collisions has led to the
novel suggestion of gluon junction transport in an initial excited
baryon and its subsequent decay into a slowly moving baryon and three
leading beam mesons \cite{junctionpapers}.
In the AMPT model, we have taken instead a more phenomenological approach 
to baryon stopping in relativistic heavy ion collisions by including 
the popcorn mechanism for baryon-antibaryon pair production from 
string fragmentation. The popcorn mechanism introduces two additional 
baryon production channels, i.e., the $B\bar B$ and $BM \bar B$ 
configurations in the Lund fragmentation model, which are controlled
by two parameters in the JETSET/PYTHIA routine \cite{Sjostrand:1993yb}
used in the HIJING model. The first parameter MSTJ(12) is changed from 
1 as set in the default HIJING model to 3 in the AMPT model
\cite{Zhang:2000bd,Lin:2001cx} in order to activate the popcorn mechanism, 
and the second parameter PARJ(5) controls the relative percentage
of the $B\bar B$ and $BM \bar B$ channels. We find that with equal 
probabilities for the $B\bar B$ and $BM \bar B$ configurations the 
net-baryon rapidity distribution at SPS can be reproduced as shown in 
Fig.~\ref{na49-dndy-newab} \cite{Lin:2001cx}. Without the popcorn 
mechanism, as in the default HIJING model, the net-baryon rapidity 
distribution would peak at a larger rapidity \cite{Zhang:1999mq}. We
also see from Fig.~\ref{na49-dndy-nofsi} that the antiproton yield
at SPS is sensitive to the antibaryon annihilation and production
channels discussed in Sec.~\ref{bbmm}. Without these reactions in 
the hadronic phase, the antiproton yield is too high compared to
preliminary NA49 data.

\subsection{Transverse momentum spectra}

\begin{figure}[ht]
\centerline{\epsfig{file=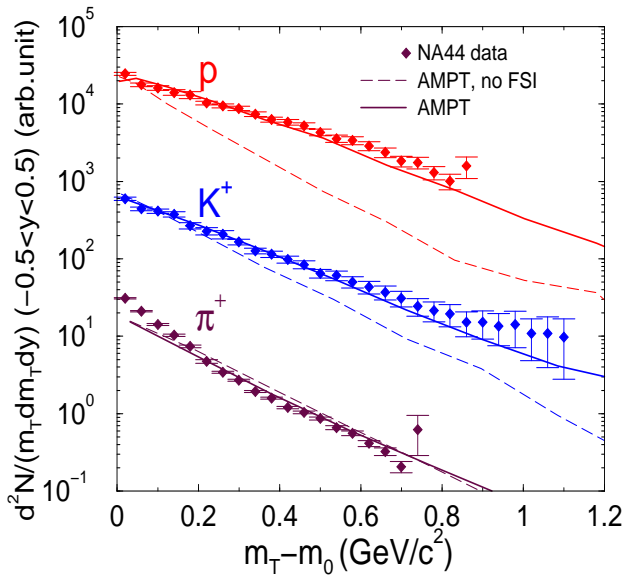,width=3.2in,height=3.2in,angle=0}}
\caption{(Color online) 
Transverse momentum spectra at mid-rapidity from the AMPT
model with and without including hadron cascade compared with experimental data
from the NA44 collaboration for central Pb+Pb collisions at SPS.}
\label{na44-mt}
\end{figure}

For the transverse momentum spectra, results from the default AMPT
model are shown by solid curves in Fig.~\ref{na44-mt} for mid-rapidity
pions, kaons, and protons in central ($b\leq 3$ fm) Pb+Pb collisions
at the SPS energy of $\sqrt{s_{NN}}=17.3$ GeV.
Compared with experimental data
from NA44, given by solid diamonds, the AMPT model gives a reasonable
description up to transverse mass of about 1 GeV$/c^2$ above the particle
mass. Without including rescatterings in the hadronic phase, the inverse 
slope parameters for the transverse mass spectra of kaons and protons 
from the AMPT model, given by dashed curves, are significantly 
reduced due to the absence of transverse flow that is induced by 
final-state interactions. 

\section{Results at RHIC energies}
\label{rhic}

\subsection{Rapidity distributions}

\begin{figure}[ht]
\centerline{\epsfig{file=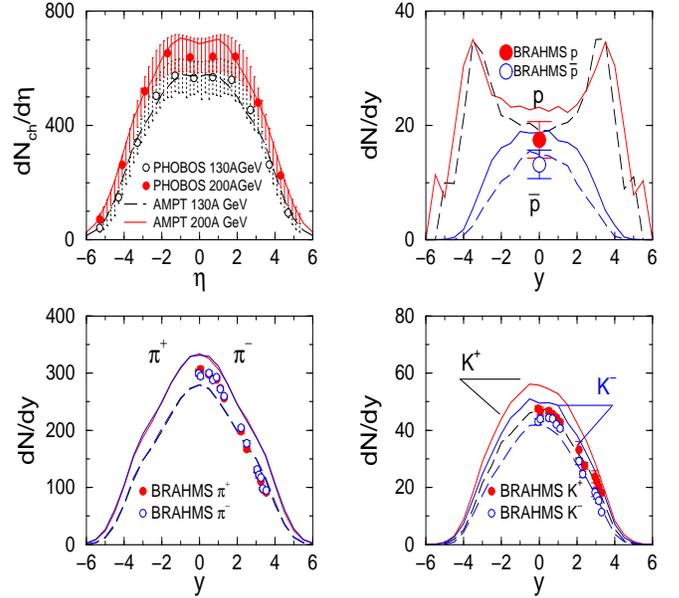,width=3.4in,height=3.2in,angle=0}}
\caption{(Color online) Rapidity distributions for Au+Au collisions at 
$\sqrt{s_{NN}}=130$ and $200$ GeV. Circles are the PHOBOS data
for 6\% most central collisions or the BRAHMS data for 5\% most
central collisions at $\sqrt{s_{NN}}=200$ GeV, and curves are results from the
default AMPT model for $b\leq 3$ fm.} 
\label{rhic-dndy}
\end{figure}

Since the number of strings associated with soft interactions in the
HIJING model depends weakly on the colliding energy, and the atomic number 
of Au is close to that of Pb, we use the same modified parameters in the Lund
string fragmentation model used for Pb+Pb collisions at SPS energies 
for Au+Au collisions at RHIC energies. In Fig.~\ref{rhic-dndy},
results for central ($b\leq 3$ fm) Au+Au collisions at center-of-mass
energies of $\sqrt{s_{NN}}=130$ GeV (dashed curves) 
and $200$ GeV (solid curves)
are shown together with data from the PHOBOS Collaboration 
\cite{Back:2000gw,Back:2001xy,phobos-new} and the BRAHMS Collaboration
\cite{Bearden:2004yx}. We find that measured total charged particle 
pseudorapidity distributions at both energies are roughly reproduced. 
More detailed comparisons on pseudorapidity distributions at different 
centralities and different RHIC energies have been carried out by the 
BRAHMS Collaboration \cite{Bearden:2001xw,Bearden:2001qq}, where 
results from the default AMPT model are compatible with the data. 
However, compared to central (top $5\%$) BRAHMS data, the AMPT model
tends to over-predict the height of the rapidity distributions 
of charged pions, kaons, protons and antiprotons. Note that the BRAHMS
data on proton and antiprotons at $y=0$ in Fig.~\ref{rhic-dndy} 
have been corrected for feed-down from weak decays. 

Other models have also been used to study hadron rapidity distributions 
at RHIC. While results from the HIJING
model \cite{Wang:xy,Wang:2000bf} are compatible with the observed
charged particle rapidity density, the model does not include
interactions among minijet partons and final-state interactions among hadrons. 
The saturation model without final-state interactions also reproduces 
the experimental data \cite{Kharzeev:2000ph,Kharzeev:2001gp}.    
Furthermore, the hadronic cascade model LUCIFER \cite{Kahana:1998wf} 
predicts a charged particle multiplicity at mid-rapidity that is 
comparable to the RHIC data \cite{Kahana:ky}. On the other hand, the
LEXUS model \cite{Jeon:1997bp}, which is based on a linear
extrapolation of ultra-relativistic nucleon-nucleon scattering 
to nucleus-nucleus collisions, predicts too many charged particles 
\cite{Jeon:2000mc} compared with the PHOBOS data. The URQMD model 
\cite{Bass:1998ca} also failed \cite{Bearden:2001xw} in describing the 
charged particle multiplicity at RHIC. 

\begin{figure}[ht]
\centerline{\epsfig{file=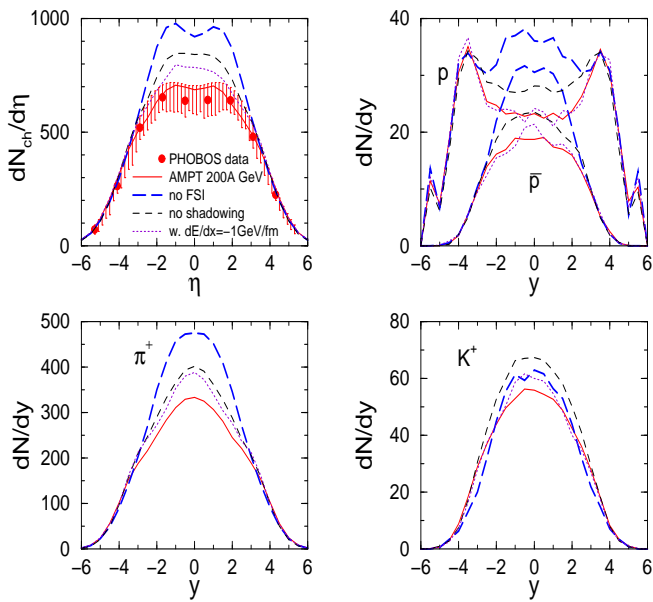,width=3.4in,
height=3.2in,angle=0}}
\caption{(Color online) 
Rapidity distributions of charged particles in central ($b\leq 3$ fm)
Au+Au collisions at $\sqrt{s_{NN}}=200$ GeV 
from the default AMPT model (solid curves), 
without final-state interactions (long-dashed curves), without nuclear
shadowing (dashed curves), and with jet quenching of $dE/dx = -1$
GeV/fm (dotted curves).} 
\label{e200-shquhadr}
\end{figure}

To see the effect of hadronic interactions, we show in 
Fig.~\ref{e200-shquhadr} by long-dashed curves the rapidity
distributions of charged particles obtained from the default AMPT model 
without final-state interactions (i.e., without parton and hadron cascades)
for central ($b\leq 3$ fm) Au+Au collisions at $\sqrt{s_{NN}}=200$ GeV. 
Compared to the case with hadronic scatterings, there is a significant 
increase in the numbers of total charged particles and pions at mid-rapidity. 
The kaon number, on the other hand, only increases slightly when
production from final-state hadronic interactions is excluded. The 
ratios of $\bar p/p$ and $K^+/\pi^+$ at mid-rapidity are $0.85$ 
and $0.13$, respectively, in the absence of final-state interactions, 
instead of $0.81$ and $0.17$ from the default AMPT model including 
the hadron cascade.  We note that although the default HIJING 
\cite{Wang:xy,Wang:2000bf} with original $a$ and $b$ parameters 
gives a total charged particle multiplicity at mid-rapidity that 
is consistent with the RHIC data, including hadronic scatterings 
reduces appreciably the final number.

We also find that excluding parton cascade in the AMPT model changes 
the final charged particle yield at mid-rapidity at 
$\sqrt {s_{NN}}=200$ GeV by less than $5\%$. This indicates that hadron 
yields are not very sensitive to parton elastic scatterings in the default
AMPT model. To take into account the effect of parton inelastic collisions
such as gluon radiation, which is not included in ZPC as it includes 
at present only elastic scatterings, we have included in the AMPT model 
the default HIJING jet quenching, i.e., an energy loss of $dE/dx = -1$ 
GeV/fm, before minijet partons enter the ZPC parton cascade. 
Results obtained with jet quenching for central Au+Au collisions
at $\sqrt{s_{NN}}=200$ GeV are shown in Fig.~\ref{e200-shquhadr} 
by dotted curves. 
We see that the quenching effects are larger for pions than 
for kaons, protons, and antiprotons. Since present calculations from 
the AMPT model without jet quenching already reproduce the data at collision 
energy of  $\sqrt{s_{NN}}=200$ GeV, 
and further inclusion of jet quenching of $dE/dx = -1$ 
GeV/fm increases the final yield of total charged particles at mid-rapidity 
by about 14\%, our results for the rapidity distribution of charged 
particles are thus consistent with a weak jet quenching at this energy. 
However, the dense matter created in heavy ion collisions expands 
rapidly, thus the energy loss at the early stage may still be large 
\cite{Wang:2001gv,Wang:2002pk}.

The effect of initial nuclear shadowing of the parton distributions in 
nuclei is also shown in Fig.~\ref{e200-shquhadr}. Without initial nuclear 
shadowing effect on minijet parton production, the charged particle 
multiplicity at mid-rapidity in central Au+Au collisions 
at $\sqrt{s_{NN}}=200$ GeV increases by about 23\%.  
Although this increase can be offset by 
using different parameters in the Lund string fragmentation, to 
reproduce both SPS and RHIC data with same parameters in the default 
AMPT model requires the inclusion of nuclear shadowing on minijet 
parton production. 

\subsection{Particle ratios}

\begin{figure}[ht]
\centerline{\epsfig{file=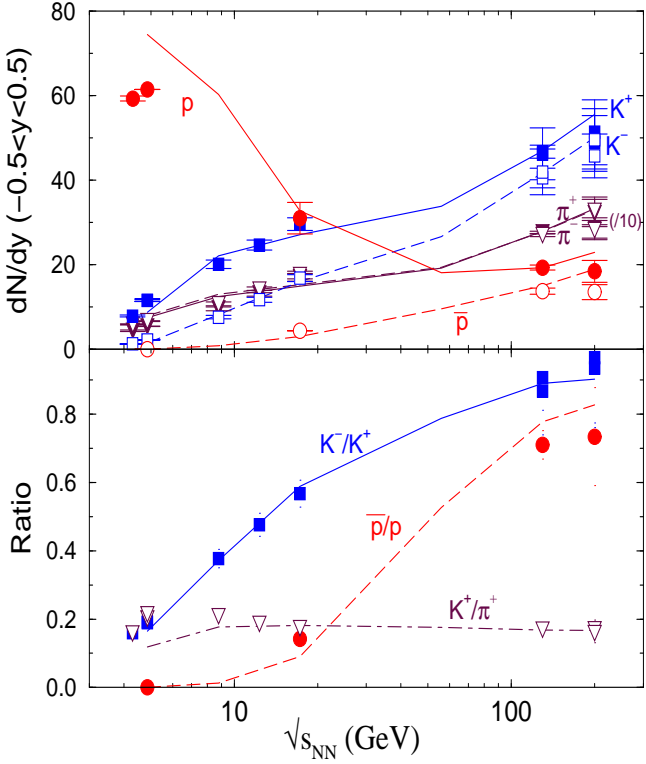,width=3.4in,height=4.in,angle=0}}
\caption{(Color online) Energy dependence of particle yields and ratios of
$K^-/K^+$, $\bar p/p$, and $K^+/\pi^+$ at mid-rapidity in central
Au+Au or Pb+Pb collisions.}
\label{ratio-e}
\end{figure}

The energy dependence of the yields of particles and their ratios at 
mid-rapidity from central Au+Au collisions at AGS to central Au+Au
collisions at RHIC energies are shown in Fig.~\ref{ratio-e}. 
Curves represent AMPT results, while data from AGS, SPS and RHIC 
experiments are shown by symbols:
triangles for pions, squares for kaons and circles for protons and antiprotons 
in the upper figure.
The AGS data are 
from the E917 and E866 Collaborations for central Au+Au collisions at 
laboratory kinetic energies of $7.94$ and $10.7$ GeV per nucleon 
\cite{e917e866}, from the E895 Collaboration for central Au+Au collisions at 
the laboratory kinetic energy of $8$ GeV per nucleon \cite{Klay:2003zf}, 
and from the E802 Collaboration for central Au+Au collisions at 
$P_{lab}=11.6-11.7$ GeV$/c$ per nucleon \cite{e802}. 
The SPS data are from NA49 for central Pb+Pb collisions at 
laboratory energies of $40A$, $80A$ and $158A$ GeV \cite{Afanasiev:2002mx}. 
The RHIC data are from the PHENIX Collaboration \cite{Adler:2003cb,phenix} 
and the STAR Collaboration \cite{Adler:2002wn,Adams:2003xp} 
for central Au+Au collisions at $\sqrt {s_{NN}}=130$ and $200$ GeV.

Since the antiproton yield increases 
with energy almost logarithmically while the proton yield initially 
decreases at low energies, the $\bar p/p$ ratio 
(data represented by circles) increases rapidly from 
almost 0 at AGS to about 0.1 at the SPS energy of $158A$ GeV 
then rapidly to about 0.8 at the
maximum RHIC energy of $\sqrt {s_{NN}}=200$ GeV, indicating the
formation of a nearly baryon-antibaryon symmetric matter at RHIC. The 
$K^+/\pi^+$ ratio (data represented by triangles), 
on the other hand, is almost constant from 
the SPS energy of $158A$ GeV to RHIC,
suggesting an approximate chemical equilibrium of strangeness in heavy 
ion collisions in this energy range \cite{kpi30}. 
The $K^-/K^+$ ratio (data represented by squares) increases 
from below 0.2 at AGS to about 0.6 at the SPS energy of $158A$ GeV, 
then gradually to about 0.9 at $\sqrt {s_{NN}}=200$ GeV, 
and the value near 1 at the top RHIC energy is closely related to 
the fact that the matter formed at RHIC is nearly baryon-antibaryon symmetric. 

\subsection{Baryon and antibaryon production}

\begin{figure}[ht]
\centerline{\epsfig{file=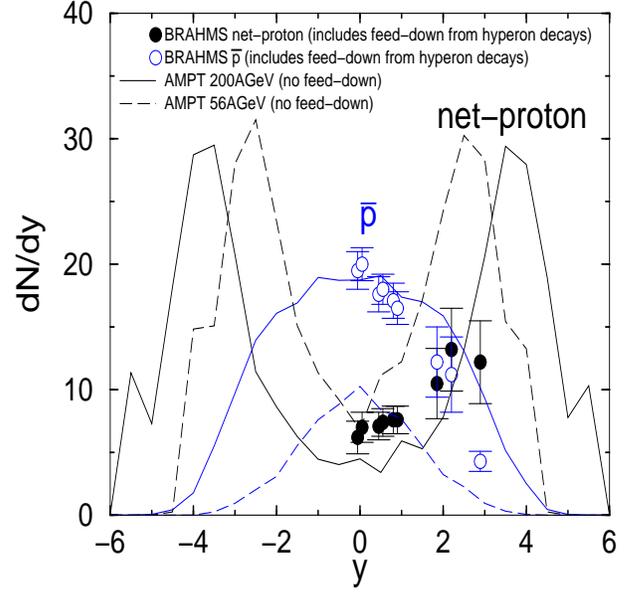,width=3.2in,height=3.2in,angle=0}}
\caption{(Color online) 
Net-proton and antiproton rapidity distributions from the 
default AMPT model for central ($b\leq 3$ fm) 
Au+Au collisions at RHIC energies of $\sqrt{s_{NN}}=200$ GeV (solid curves) 
and $56$ GeV (dashed curves) versus data 
from the BRAHMS Collaboration for $5\%$ most central collisions at 
$\sqrt{s_{NN}}=200$ GeV.}
\label{rhic-netppbar}
\end{figure}

Fig.~\ref{rhic-netppbar} shows the net-proton and antiproton rapidity 
distributions from the default AMPT model at two different RHIC energies.   
Filled and open circles are BRAHMS data on net-proton and 
antiprotons without taking into account corrections due to weak decay
of hyperons \cite{Bearden:2003hx}, i.e., the data include feed-down 
from hyperon weak decays. Since the AMPT results on proton and
antiprotons exclude feed-down from weak decays, the antiproton yield
at mid-rapidity is actually over-predicted by the AMPT model, 
as can be more clearly seen in Fig.~\ref{rhic-dndy}; while the net-proton 
rapidity distributions may agree reasonably with the BRAHMS data.

It is further seen that the antiproton yield at mid-rapidity increases 
rapidly with collision energy, and peaks of net-proton and proton 
distributions shift toward larger rapidity at higher collision energies. 
Since the proton yield at mid-rapidity first decreases from the AGS 
energy to the RHIC energy of $\sqrt{s_{NN}}=56$ GeV 
and eventually increases slowly with energy (see Fig.~\ref{ratio-e}), 
it may have a minimum between the SPS energy and
the maximum RHIC energy. On the other hand, the net-proton yield at 
mid-rapidity decreases with the colliding energy in the energy range
from the AGS energy to the maximum RHIC energy. 

\subsection{Transverse momentum spectra}

\begin{figure}[ht]
\centerline{\epsfig{file=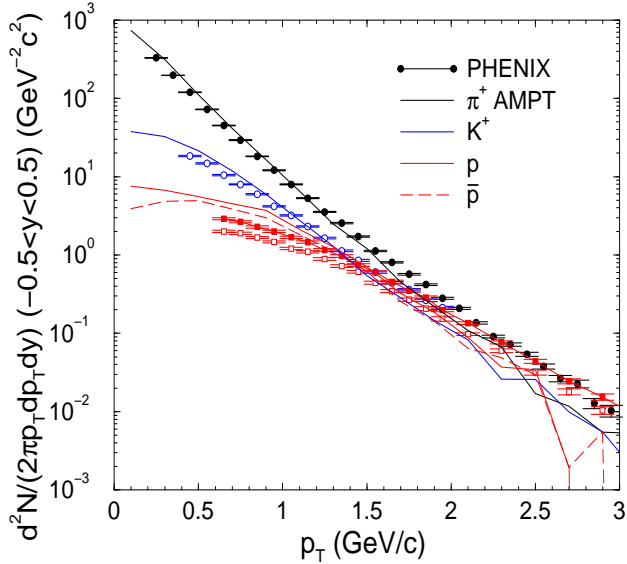,width=3.2in,height=3.2in,angle=0}}
\caption{(Color online) 
Transverse momentum spectra of mid-rapidity pions, kaons,
protons and antiprotons from central ($b\leq 3$ fm) Au+Au collisions at
$\sqrt{s_{NN}}=200$ GeV from the default AMPT model.  Data are from the PHENIX
Collaboration.}
\label{e200-pt}
\end{figure}

For transverse momentum spectra, results from the default AMPT model 
for pions, kaons, and protons at mid-rapidity from central ($b \leq 3$
fm) Au+Au collisions at $\sqrt{s_{NN}}=200$ GeV are shown in Fig.~\ref{e200-pt}
together with the 5\% most central data from the PHENIX Collaboration 
\cite{Adler:2003cb}. Below $\pt=2$ GeV/$c$, the default AMPT model
gives a reasonable description of the pion and kaon spectra 
\cite{Lin:2001yd}. It over-predicts, however, both the
proton and antiproton spectra at low $\pt$. We note that the PHENIX
data have been corrected for weak decays.

\begin{figure}[ht]
\centerline{\epsfig{file=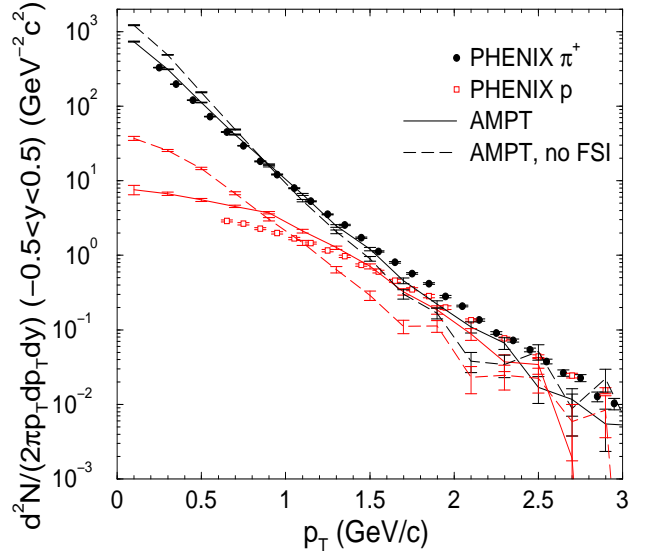,width=3.2in,height=3.2in,angle=0}}
\caption{(Color online) 
Transverse momentum spectra of mid-rapidity pions and protons 
from the default AMPT model with (solid curves) and without (dashed curves) 
final-state interactions in central Au+Au collisions at $\sqrt{s_{NN}}=200$ 
GeV.} 
\label{e200-pt-fsi}
\end{figure}

The effect due to final-state hadronic scatterings on pion and proton 
transverse momentum spectra is illustrated in Fig.~\ref{e200-pt-fsi},
where solid and dashed curves with statistical errors
are, respectively, the results with and without hadron cascade
in the default AMPT model. As found in Fig.~\ref{na44-mt} at SPS
energies, hadronic rescatterings increase significantly the inverse
slope of the proton transverse momentum spectrum while they do not
affect much that of pions. As a result, the final proton yield at 
mid-rapidity becomes close to the pion yield at $\pt \sim 2$ GeV/$c$,
as observed in experiments at RHIC \cite{Adcox:2001mf,Adler:2003cb}. 
Without final-state scatterings, the proton yield given by the dashed
curve is well below the pion yield up to $\pt \sim 1.7$ GeV/$c$ when
statistical fluctuations in the AMPT calculations become large.  Results
from the default AMPT model thus indicates that the observed large 
$p/\pi$ ratio at $\pt \sim 2$ GeV/$c$ is due to the collective transverse 
flow generated by final-state rescatterings. 

\begin{figure}[ht]
\centerline{\epsfig{file=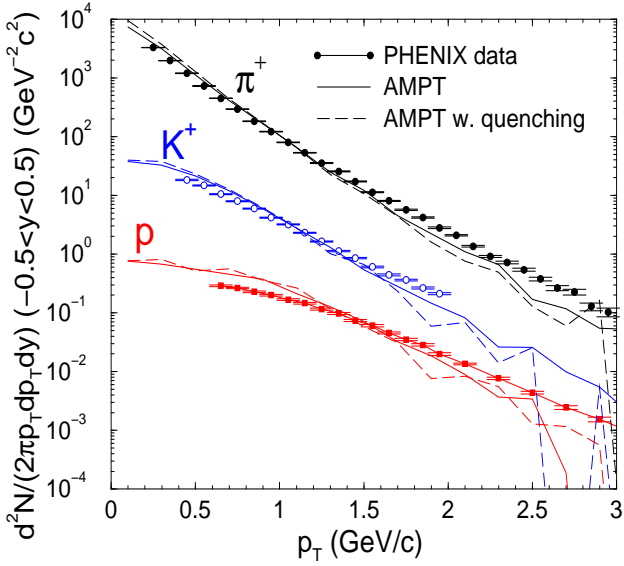,width=3.2in,
height=3.2in,angle=0}}
\caption{(Color online) Transverse momentum spectra 
with and without quenching of $dE/dx=-1$ GeV/fm  
for central Au+Au collisions at $\sqrt{s_{NN}}=200$ GeV from the default 
AMPT model versus data from the PHENIX Collaboration.}
\label{e200-pt-quenching}
\end{figure}

\begin{figure}[ht]
\centerline{\epsfig{file=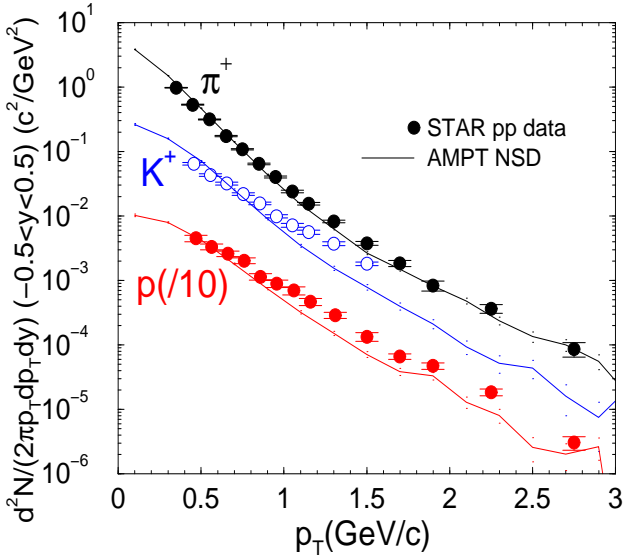,width=3.2in,
height=3.2in,angle=0}}
\caption{(Color online) 
Transverse momentum spectra of mid-rapidity pions, kaons 
and protons for $pp$ collisions at $200$ GeV from the default 
AMPT model versus data from the STAR Collaboration.}
\label{e200-pt-pp}
\end{figure}

The above results are obtained without quenching of minijet partons due 
to gluon radiations. Including this effect would reduce the yield of 
hadrons at large transverse momentum as shown in 
Fig.~\ref{e200-pt-quenching}, leading to an appreciable
discrepancy between the AMPT results and the experimental data 
from the PHENIX collaboration for hadrons with momenta greater than 
1.5 GeV$/c$.  This discrepancy may stem, however, from the fact that
the AMPT model already underpredicts the STAR data for hadron 
yield above 1 GeV$/c$ in $pp$ collisions as shown in Fig.~\ref{e200-pt-pp}.  
Since the STAR results are consistent with previous UA5/CDF 
measurements at similar multiplicities \cite{Adams:2003xp} but have
significantly higher mean kaon and proton transverse momenta 
than interpolated values from the UA5/CDF $pp$/$p\bar p$ data 
shown in Fig.~\ref{meanpt-s}, the STAR NSD trigger might be quite 
different from the UA5/CDF NSD trigger. To make a more definitive 
conclusion on jet quenching in AA collisions thus requires a 
focused and higher-statistics study of $\pt$ spectra in both 
$pp$ and Au+Au collisions.

\subsection{Phi meson yield via dilepton and dikaon spectra}

Since the $\phi$ meson is unstable, it can only be detected from its 
decay products such as the kaon-antikaon pair or the lepton pair.
In the AMPT model, we follow the production and decay of $\phi$ mesons 
as well as the scattering of the kaon daughters from their decays. Since the
latter destroys the possibility of reconstructing the parent $\phi$ 
mesons, only $\phi$ mesons with their decay kaons not undergoing scattering
can be reconstructed from the kaon-antikaon invariant mass distribution.
The kaon-antikaon pair from a $\phi$ meson decay is likely to
undergo rescatterings in the medium, which would lead to a
reconstructed invariant mass outside the original $\phi$ meson peak. 
Hence $\phi$ mesons decaying in the dense medium are difficult 
to be identified via reconstructed kaon-antikaon pairs. In contrast, 
dileptons have negligible final-state interactions with the surrounding 
hadronic medium, they are thus considered to carry useful information
about hadron properties in hot dense hadronic matter 
\cite{Li:1994cj,Chung:1998ev}, which are expected to be different from
those in free space \cite{Asakawa:1994tp,Ko:1997kb}.

Since dimuons are emitted continuously during evolution of the system,
the total number of dimuons is given by
\ber
N_{\mu^+\mu^-} &=& \int_0^{t_{cut}} \! dt \ N_\phi(t) 
\Gamma_{\phi\to \mu^+\mu^-}(M) \nonumber \\
&+& N_\phi(t_{cut}) \frac{\Gamma_{\phi\to \mu^+\mu^-}}{\Gamma_\phi} ~,
\label{muphi}
\eer
where $N_\phi(t)$ denotes the number of $\phi$ mesons at time $t$ and
$\Gamma_{\phi\to \mu^+\mu^-}(M)$ is its decay width to 
dimuon as given by
\ber
\Gamma_{\phi\to \mu^+\mu^-}\! ( \! M \!) \!=\! C_{\mu^+\mu^-}
\frac{m^{4}_\phi}{M^3} \! 
\left( \! 1 \!-\! \frac{4m^2_\mu}{M^2} \! \right)^{\! 1/2}
\!\!\! \left( \! 1 \!+\! \frac{2m^2_\mu}{M^2} \! \right) . 
\label{gamamu}
\eer
The coefficient 
$C_{\mu^+\mu^-} \equiv \alpha^2 \big/27(g^2_{\phi K\bar K}/4\pi) 
= 1.634\times 10^{-6}$ is determined from measured width 
of free $\phi$ meson at invariant mass $M=m_\phi$. 
In Eq.~(\ref{muphi}), the first term on the right-hand-side
corresponds to dimuon production before a cut-off time $t_{cut}$ 
during the hadronic phase of heavy ion collisions, while the
second term refers to dimuon emission after that cut-off time. 
The reconstructed $\phi$ meson number is then obtained by dividing the 
above expression by the dimuon branching ratio in free-space of  
$\Gamma_{\phi\to \mu^+\mu^-}/\Gamma_\phi = 3.7\times 10^{-4}$. 
The number of kaon-antikaon pairs coming from $\phi$ meson decays can be 
similarly expressed as Eq.~(\ref{muphi}):
\ber
N_{K\!\bar K} \!\!=\!\! \int_0^{t_{cut}} \!\!\! dt \ N_\phi(t) 
\Gamma_{\phi\to K\bar K}
(M) \!+\! N_\phi(t_{cut}) \frac{\Gamma_{\phi\to K\bar K}}{\Gamma_\phi} .
\label{kkphi}
\eer
The $\phi$ meson number from kaon-antikaon decays is then obtained by 
dividing Eq.~(\ref{kkphi}) by the $K\bar K$ branching ratio in free-space
$\Gamma_{\phi\to K\bar K}/\Gamma_\phi$. 

Using the default AMPT model, we have studied the $\phi$ meson yield 
reconstructed from $K^+K^-$ and $\mu^+\mu^-$ pairs for central Pb+Pb
collisions at SPS and for central Au+Au collisions at RHIC \cite{Pal:2002aw}. 
In both cases we have found that, due to hadronic rescattering and
absorption in the kaonic channel, the $\phi$ meson abundance at
mid-rapidity is larger in the dimuon channel than in the dikaon channel, 
and the inverse slope parameter obtained from the transverse mass
spectra of $\phi$ mesons in the range $0<m_T-m_\phi<1$ GeV$/c^2$ is larger
in the dimuon channel than in the dikaon channel. These features are
consistent with the data at SPS from the NA49 \cite{Afanasev:2000uu} 
and NA50 \cite{Abreu:2001qp} collaborations. Comparison of the results for RHIC
with future experimental data will allow us to learn if enhanced $\phi$ meson 
production is due to the formation of the quark-gluon plasma during
the early stage of collisions \cite{Pal:2002aw}. 

\subsection{Multistrange baryon production}

One possible signal for the quark-gluon plasma is enhanced production
of strange particles \cite{Muller:1980kf}, particularly those consisting of 
multistrange quarks such as $\Xi$ and $\Omega$ baryons as well as their 
antiparticles. The argument is that the rate for strange hadron 
production is small in hadronic matter due to large threshold 
and small cross sections while the production
rate for strange quarks is large in the quark-gluon plasma
\cite{Koch:1986ud}.

A detailed study of multi-strange baryons in the AMPT model \cite{Pal:2001zw} 
shows that, although few multi-strange baryons are produced in HIJING 
(see Table I of Ref.~\cite{Pal:2001zw}), 
including the strangeness-exchange interactions listed in Sec.~\ref{ms}
in the default AMPT model leads to an enhanced production of
multistrange baryons in heavy ion collisions at both SPS and RHIC.
For central Pb+Pb collisions at SPS, the strangeness-exchange reactions 
enhance modestly the yields of $\Lambda$, $\Sigma$, and $\Xi$, but 
increase the $\Omega$ yield by more than an order of magnitude. 
However, the $\Omega$ yield in central Pb+Pb collisions at SPS 
from the default AMPT model is still about a factor
of 2 lower than the data, and this may indicate that strangeness
production is already enhanced during early stage of collisions. 

Predictions from the default AMPT model \cite{Pal:2001zw} also show
that the slope parameters obtained from the transverse mass spectra of 
multistrange baryons reveal a plateau structure since these particles, 
mostly generated by strangeness-exchange reactions in the model, are weakly 
interacting and decouples rather early from the system. It will be
very interesting to test these predictions against the RHIC data. 
In particular, the quark coalescence model 
\cite{Lin:2002rw,Voloshin:2002wa,Molnar:2003ff,Fries:2003vb,Greco:2003xt,Lin:2003jy,Greco:2003vf,Greco:2003mm,Fries:2003kq},
which assumes that all energy in the early stage of RHIC collisions is
in the partonic degrees of freedom as assumed in the string melting scenario
\cite{Lin:2001zk,Lin:2002gc,Lin:2003iq} of the AMPT model, 
relates the transverse momentum spectra or the elliptic flow of 
multistrange baryons with other hadrons such as kaons and protons.  
These relations can be drastically different from the default AMPT 
model, where the partonic stage includes only minijet partons and 
has a much smaller effect than later hadronic stage.

\subsection{Equilibration between $J/\psi$ and open charm}

$J/\psi$ production has long been proposed as a possible signal for
the formation of the quark-gluon plasma in relativistic heavy ion 
collisions \cite{Matsui:1986dk}.  Based
on the expectation that the effective potential between charm and
anticharm quarks changes in the QGP due to color screening effect,
they will not form bound states above certain critical temperature,
which is somewhat higher than the deconfinement temperature. The
qualitative change in the heavy quark effective potential has been
verified by results from lattice QCD simulations 
\cite{Kaczmarek:1999mm,Karsch:2001vs,Bornyakov:2002iv}. As a result, 
$J/\psi$ production is expected to be suppressed if the QGP is formed in 
heavy ion collisions. In particular, the observed abnormal suppression 
of $J/\psi$ in central Pb+Pb collisions at SPS has been attributed to 
the formation of the quark-gluon plasma in these collisions 
\cite{Kharzeev:1996yx,Heinz:2000bk}. However, there are other possible 
mechanisms for $J/\psi$ suppression in heavy ion collisions 
\cite{Alde:1990wa,Leitch:1999ea,Abreu:jh}, e.g., $J/\psi$ may be
destroyed by collisions with incoming nucleons or with gluons in the
initial partonic matter \cite{Shuryak:1978ij,Xu:1995eb} or with
comoving hadrons in the hadronic matter 
\cite{Gavin:1996yd,Cassing:1996zb,Geiss:1998ma,Kahana:1998cw,Spieles:1999kp,Sa:mi,Spieles:pm,Capella:2000zp,Haglin:1999xs,Lin:1999ad,Liu:2001ce,Sibirtsev:2000aw,Navarra:2001jy,Wong:1999zb,Wong:2001an}.

For heavy ion collisions at RHIC energies, multiple pairs of charm-anticharm 
quarks can be produced in one event. Estimates using the kinetic approach 
\cite{Thews:2000rj} have shown that the number of $J/\psi$ produced from 
the interaction between charm and anticharm quarks, i.e., the inverse
reaction of $J/\psi$ dissociation by gluons, may exceed that expected from 
the superposition of initial nucleon-nucleon interactions. Calculations 
based on the statistical model 
\cite{Braun-Munzinger:2000px,Braun-Munzinger:2000ep,Grandchamp:2001pf},
which assumes that $J/\psi$ formed during hadronization of the QGP
is in chemical equilibrium with charm mesons, also predict that
the $J/\psi$ number is comparable to the expected primary yield.

To study the above discussed effects on $J/\psi$ production, the default 
AMPT model has been modified \cite{amptjpsi} to include $J/\psi$ 
absorption and production in both initial partonic and final hadronic 
matters \cite{Zhang:2000nc}.  It was found that because of these final-state 
interactions the final $J/\psi$ yield in central Au+Au collisions at RHIC 
may exceed the suppressed primary $J/\psi$ yield 
which is based only on the color screening mechanism. 
Furthermore, the $J/\psi$ yield at RHIC depends on the charm quark mass 
in the partonic matter and charmed hadron masses in the hadronic matter. 

We note that the elliptic flows of $J/\psi$ and charmed hadrons 
are related in the quark coalescence model, and they exhibit novel
mass effects when the constituent quark masses in the hadron are
different \cite{Lin:2003jy}. The transverse momentum spectra of
electrons from open charm decays are, however, insensitive to the charm
transverse flow \cite{Batsouli:2002qf}. Comparing RHIC data on charm
hadron or electron $v_2$ \cite{Greco:2003vf} with results from the quark
coalescence model may allow us to determine whether in heavy ion 
collisions charm quarks flow collectively as light quarks.

\subsection{Elliptic flow}
\label{v2}

Elliptic flow in heavy ion collisions is a measure of the asymmetry of 
particle momentum distributions in the transverse plane and is
generated by the anisotropic pressure gradient in initial hot dense
matter as a result of the spatial asymmetry in non-central collisions
\cite{Barrette:1994xr,Appelshauser:1998dg,Ollitrault:1992bk,Rqmd,Danielewicz:1998vz,Zheng:1999gt}.
It is defined as one half of the second Fourier coefficient of the
azimuthal angle distribution of particle transverse momentum and can
be evaluated as 
\begin{equation}
v_2 =\left \langle {{p_x^2-p_y^2} \over {p_x^2+p_y^2}} \right \rangle. 
\end{equation}
In the above, the $x$-axis is along the impact parameter in the
transverse plane of each event while the $y$-axis points out of the
reaction plane, and the average is taken over particles in consideration. 

At RHIC, large hadron elliptic flows have been observed  
\cite{Ackermann:2001tr,Lacey:2001va,Snellings:2001nf,Park:2001gm,Back:2002gz,Adler:2002pu,Adcox:2002ms},
and their dependence on transverse momentum 
\cite{Ackermann:2001tr,Lacey:2001va}
and pseudorapidity \cite{Roland:2001me,Park:2001gm} as well as on
particle species \cite{Snellings:2001nf,Adler:2001nb} have also been
studied. To understand these experimental results, many theoretical
models have been introduced, and these include semi-analytic models
\cite{Heiselberg:1999mf}, models based on parton energy loss 
\cite{Gyulassy:2001gk,Wang:2001fq}, 
hydrodynamic models \cite{Ollitrault:1992bk,Heiselberg:1999es,Huovinen:2001cy,Huovinen:2001wn}, transport models 
\cite{Rqmd,Zabrodin:2001rz,Zhang:1999rs,Molnar:v2,Lin:2001zk,Humanic:2002iw},
including a hybrid model that uses the transport model as an
after-burner of the hydrodynamic model \cite{Teaney:v2}, and quark
coalescence models \cite{Lin:2002rw,Voloshin:2002wa,Molnar:2003ff,Fries:2003vb,Greco:2003xt,Lin:2003jy,Greco:2003vf,Greco:2003mm,Fries:2003kq}.
Transport models based on hadronic and/or string degrees of freedom in 
general give a smaller elliptic flow \cite{Snellings:2001nf} than that
observed at RHIC.  Although hydrodynamic models can explain the large
elliptic flow at low transverse momenta with a large initial energy
density, they predict a continuously increasing elliptic flow with
increasing hadron transverse momenta instead of the observed level
off in experiments
\cite{Snellings:2001nf,Adler:2002pu,Adler:2002pb,Adler:2002ct},
indicating that high transverse momentum particles do not reach thermal
equilibrium. With the AMPT model, we can address such non-equilibrium 
effects to obtain information on the degree of thermalization in these
collisions. It further allows us to study the effects due to both
partonic and hadronic rescatterings. 

\begin{figure}[ht]
\centerline{\epsfig{file=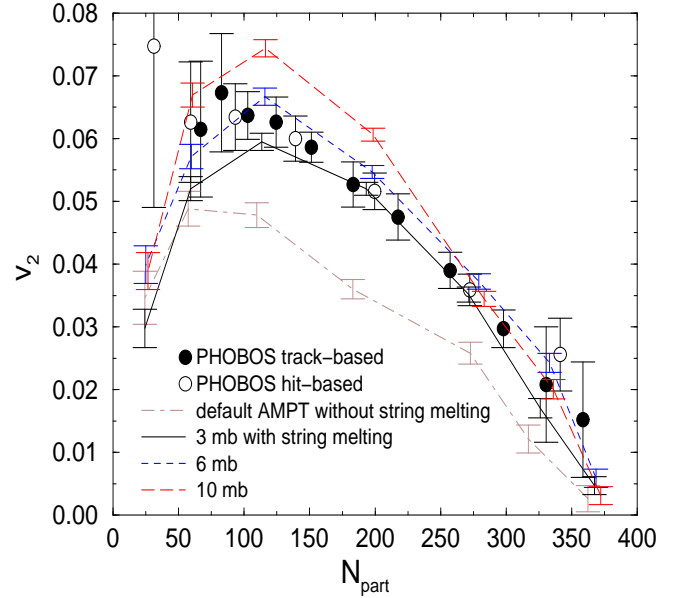,width=3.4in,height=3.2in,angle=0}}
\caption{(Color online) Centrality dependence of charged hadron elliptic flow 
in Au+Au collisions at $\sqrt {s_{NN}}=200$ GeV. Data from the PHOBOS
collaboration \protect\cite{phobos-200v2} are shown by circles, while 
results from the AMPT mode with different partonic dynamics are shown
by curves.}
\label{v2-b}
\end{figure}

In Fig.~\ref{v2-b}, we show the elliptic flow of all charged particles 
in Au+Au collisions at $\sqrt {s_{NN}}=200$ GeV from both default
AMPT model and the extended model with string melting as a function of
$N_{\rm part}$, which is the total number of participant nucleons
after primary collisions but before partonic and hadronic
rescatterings \cite{Lin:2003ah}. To compare with PHOBOS
data \cite{phobos-200v2}, we have included only charged particles 
with pseudorapidities and transverse momenta in the ranges
$\eta \in (-1, 1)$ and $\pt \in (0.1, 4)$ GeV/$c$, respectively,
for evaluating the elliptic flow. Error bars in the figure
represent the statistical error in our calculations. We see that
the value of $v_2$ depends not only on whether initial strings 
are converted to partons but also on the parton cross section used in
the model, similar to that seen in heavy ion collisions at 
$\sqrt{s_{NN}}=130$ GeV \cite{Lin:2001zk}. 
With the default AMPT model, which includes only 
scattering among minijet partons, the elliptic flow obtained with
a parton scattering cross section of 3 mb is found to be too small
to account for the experimental data from the PHOBOS collaboration.  
The elliptic flow is larger in the string melting scenario as
soft partons from melted strings also participate in partonic
scatterings, and its value is further increased when the parton
scattering cross section increases from 3 mb to 10 mb, indicating that
the dense system created in heavy ion collisions at RHIC does not
quite reach thermal equilibrium. 

\begin{figure}[ht]
\centerline{\epsfig{file=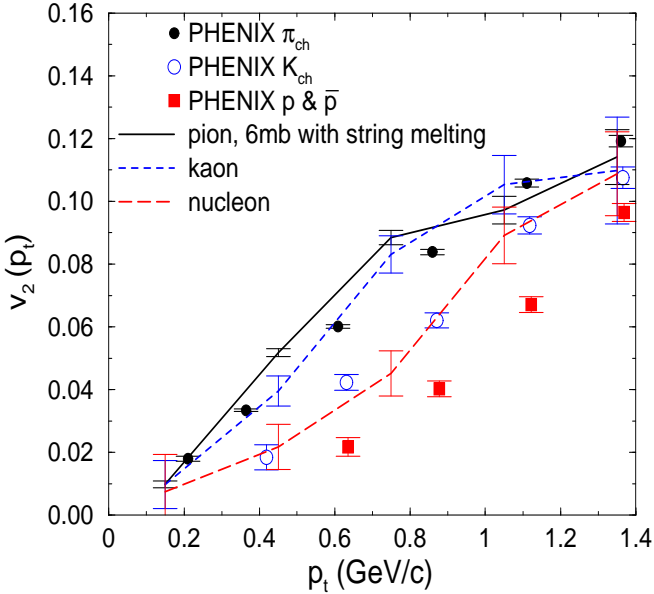,width=3.4in,
height=3.2in,angle=0}}
\caption{(Color online) 
Transverse momentum dependence of elliptic flow of identified
hadrons for minimum-bias Au+Au collisions at $\sqrt{s_{NN}}=200$ GeV. 
Circles are 
data from the PHENIX Collaboration \protect\cite{Adler:2003kt} while
curves are results from the AMPT model with string melting and using
6 mb for the parton scattering cross section.}
\label{v2-pt}
\end{figure}

In Figure~\ref{v2-pt}, the differential elliptic flows $v_2(\pt)$ of
pions, kaons and protons from the AMPT model in the string melting
scenario with a parton scattering cross section of 6 mb are shown 
for minimum-bias Au+Au collisions at $\sqrt{s_{NN}}=200$ GeV. 
Although the PHENIX
data are for particles with pseudorapidities $\eta \in (-0.35, 0.35)$, 
the AMPT results are for $\eta \in (-1, 1)$ in order to improve the
statistics of our calculation. In this low $\pt$ region, we observe
the mass ordering of $v_2(\pt)$, i.e., hadrons with larger masses have
smaller $v_2$ values at the same $\pt$, similar to that from the
hydrodynamic model \cite{Kolb:2000fh,Huovinen:2001cy}. For hadrons 
at higher $\pt$, which are not studied here, a scaling of hadron
elliptic flows according to their constituent quark content has been 
proposed based on the quark coalescence model 
\cite{Lin:2002rw,Voloshin:2002wa,Molnar:2003ff,Kolb:2004gi}. 
Except for pions, this scaling seems to be confirmed by experimental 
data from both STAR \cite{Sorensen:2003wi} and PHENIX
\cite{Adler:2003kt} collaborations. The violation of pion elliptic
flow from the constituent quark number scaling has been attributed to 
effects due to resonance decays \cite{Greco:2004ex,Dong:2004ve} 
and the binding energy of hadrons \cite{Lin:2003jy}. 
 
The pseudorapidity dependence of the elliptic flow in minimum-bias 
Au+Au collisions at $\sqrt{s_{NN}}=200$ GeV 
has also been studied in the AMPT model
\cite{Chen:2004vh}. While the string melting scenario with 
a parton scattering cross section of 3 mb as well as the default AMPT model 
gives a better description of the PHOBOS data \cite{phobos-200v2} 
at large pseudorapidity, the string melting scenario 
with a larger parton scattering cross section reproduces the PHOBOS data 
at mid-pseudorapidity. This may not be 
unreasonable as not all strings are expected to melt at large 
pseudorapidity where the smaller particle multiplicity leads to 
a lower energy density.

The AMPT model has further been used to study higher-order anisotropic 
flows $v_{4}$ and $v_{6}$ of charged hadrons at mid-rapidity in 
heavy ion collisions at RHIC \cite{Chen:2004dv}. It was found that the
same large parton scattering cross section used in explaining the measured 
$v_2$ of charged hadrons could also reproduce recent data on their 
$v_{4}$ and $v_{6}$ from Au+Au collisions at $\sqrt{s_{NN}}=200$ GeV 
\cite{Adams:2003zg}. Furthermore, the $v_{4}$ was seen to be a more 
sensitive probe of the initial partonic dynamics in these collisions than
$v_{2}$. Moreover, higher-order parton anisotropic flows are
non-negligible and satisfy the scaling relation $v_{n,q}(\pt)\sim
v_{2,q}^{n/2}(\pt)$, which leads naturally to the observed similar scaling
relation among hadron anisotropic flows when the coalescence model is used
to describe hadron production from the partonic matter.

\subsection{Two-particle interferometry}
\label{hbt}

The Hanbury-Brown Twiss (HBT) effect was first used to measure the
size of an emission source like a star \cite{hbt}. For heavy ion
collisions, HBT may provide information not only on the spatial 
extent of the emission source but also on its emission duration 
\cite{Pratt:su,Bertsch:1988db,Pratt:zq,Rischke:1996em}. In particular, 
the long emission time as a result of the phase transition from the 
quark-gluon plasma to the hadronic matter in relativistic heavy ion collisions 
may lead to an emission source which has a $R_{\rm out}/R_{\rm side}$
ratio much larger than one \cite{Rischke:1996em,Soff:2000eh,Soff:2001hc}, 
where the out-direction is along the total transverse momentum 
of detected two particles and the side-direction is perpendicular to
both the out-direction and the beam direction (called the long-direction)  
\cite{Bertsch:1988db,Pratt:zq}. Since the quark-gluon plasma is
expected to be formed in heavy ion collisions at RHIC, it is thus
surprising to find that the extracted ratio $R_{\rm out}/R_{\rm side}$
from a Gaussian fit to the measured correlation function of two
identical pions in Au+Au collisions is close to one 
\cite{Adler:2001zd,Johnson:2001zi,Adcox:2002uc}. 
This is in sharp contrast with calculations from hydrodynamic models  
\cite{Soff:2000eh,Heinz:2002un}, where $R_{\rm out}/R_{\rm side}$ is
typically well above one.

Denoting the single-particle emission function by $S(x,{\bf p})$, the
HBT correlation function for two identical bosons in the absence of
final-state interactions is given by 
\cite{Pratt:su,Wiedemann:1999qn}:
\ber
&&C_2({\bf Q},{\bf K})=1+ \nonumber \\
&&\frac{\int \! d^4x_1d^4x_2 S(x_1,\! {\bf K})
S(x_2,\! {\bf K}) \cos \left [Q \! \cdot \! (x_1-x_2) \right ]}
{\int d^4x_1 S(x_1,{\bf p_1}) \int d^4x_2 S(x_2,{\bf p_2})},
\label{emission}
\eer
where ${\bf p_1}$ and ${\bf p_2}$ are momenta of the two hadrons, 
${\bf K}=({\bf p_1}+{\bf p_2})/2$, and 
$Q=({\bf p_1}-{\bf p_2}, E_1-E_2)$.  The three-dimensional correlation
function in ${\bf Q}$ can be shown as one-dimensional
functions of the projections of ${\bf Q}$ in the ``out-side-long''
coordinate system \cite{Bertsch:1988db,Pratt:zq}.

Expecting that the emission function is sufficiently smooth in momentum  
space, the size of the emission source can then be related to the
emission function as:
\begin{eqnarray}
&&R_{ij}^2(K)=\left . -{1 \over 2}
\frac{\partial^2 C_2({\bf Q}, {\bf K})}{\partial Q_i \partial Q_j} 
\right |_{{\bf Q}=0} \nonumber \\
&\! =\! & 
\! \! D_{x_i,x_j}\! (\! K\! )\! -\! D_{x_i,\beta_j t}\! (\! K\! )\! -\! 
D_{\beta_i t,x_j}\! (\! K\! ) \! +\! D_{\beta_i t,\beta_j t}\! (\! K\! ). 
\label{source}
\end{eqnarray}
These source radii are thus expressed in terms of the space-time variances, 
$D_{x,y}=\langle x\cdot y \rangle-\langle x\rangle \langle y \rangle$, 
with $\langle x\rangle$ denoting the average value of $x$.  
In the above, $x_i$'s $(i=1-3)$ denote the projections of the particle
position at freeze-out in the ``out-side-long'' system, 
i.e., $x_{\rm out}$, $ x_{\rm side}$ and $x_{\rm long}$, respectively; 
and ${\bf \beta}={\bf K}/K_0$ with $K_0$ being the average energy of
the two particles.  

The experimentally measured two-particle correlation function 
$C_2({\bf Q},{\bf K})$ in central heavy ion collisions is usually fitted 
by a four-parameter Gaussian function after correcting for final-state 
Coulomb interactions, i.e., 
\begin{eqnarray}
C_2({\bf Q},{\bf K})=1+ \lambda \exp 
\left ( -\sum_{i=1}^3 R^2_{ii}(K) Q_i^2 \right ).  
\label{hbt-fit}
\end{eqnarray}
If the emission source is Gaussian in space-time, the fitted radii
$R_{ii}$ would be identical to the source radii determined from the
emission function via Eq.~(\ref{source}). However, because of
space-time correlations in the emission function, such as those
induced by the collective flow, the fitted radii can be quite
different from the source radii \cite{Sullivan:wb,Hardtke:1999vf}. 

We have used the AMPT model to study the interferometry of two
identical pions or kaons in central Au+Au collisions at RHIC 
\cite{Lin:2002gc,Lin:2003iq}.  The source of the emitted particles
have been obtained from their space-time coordinates $x$ and momenta 
${\bf p}$ at kinetic freeze-out, i.e., at their last interactions. 
The correlation function $C_2({\bf Q},{\bf K})$ is then evaluated 
in the frame of longitudinally comoving system using the program
Correlation After Burner \cite{pratt:uf}. In these calculations, 
the cut-off time $t_{cut}$ for hadron cascade has been chosen as 
200 fm$/c$ for HBT studies instead of the default 30 fm$/c$ as
we are interested in the space-time and momentum distributions of 
hadrons at freeze-out even though their rapidity distributions and
momentum spectra essentially do not change after 30 fm$/c$.

\begin{figure}[ht]
\centerline{\epsfig{file=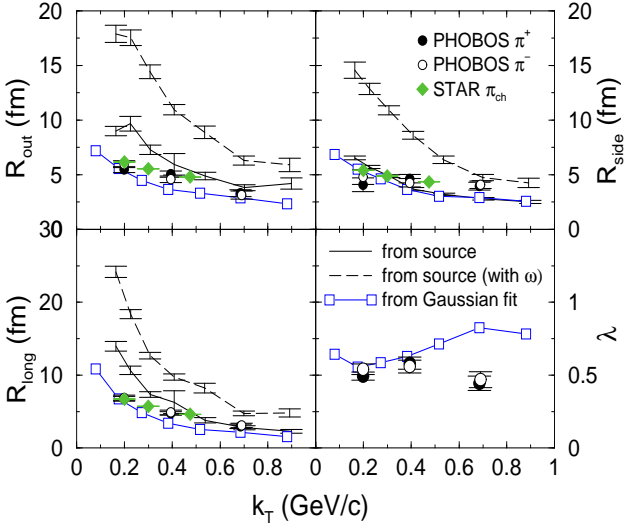,width=3.3in,
height=2.8in,angle=0}}
\caption{(Color online) 
Pion source radii with (dashed curves) and without (solid curves) 
$\omega$ decays as well as the fitted radius and $\lambda$ parameters
from the Gaussian fit to the correlation function (curves with
squares) at mid-rapidity as functions of $k_{\rm T}$ from AMPT with
string melting and $\sigma_p=6$mb for central Au+Au collisions at 
$\sqrt{s_{NN}}=200$ GeV. 
Circles are the PHOBOS data, while diamonds are the STAR data.}
\label{rkt}
\end{figure}

For central ($b=0$ fm) Au+Au collisions at $\sqrt {s_{NN}}=200$ GeV,
results for charged pions from the AMPT model with string melting
and parton cross section $\sigma_p=6$ mb are shown in Fig.~\ref{rkt} 
together with experimental data from the STAR \cite{Adams:2003ra} 
and PHOBOS \cite{Back:2004ug} Collaborations. The STAR data at the
three transverse momentum $k_{\rm T}$ bins of $(0.15, 0.25)$,
$(0.25, 0.35)$, and $(0.35, 0.60)$ GeV$/c$ are for mid-rapidity and 
0-5\% most central collisions, while the PHOBOS data are for 0-15\%
most central collisions with $\langle y_{\pi\pi} \rangle=0.9$. We find
that the source radii $R_{\rm out}$ (upper-left panel), $R_{\rm side}$ 
(upper-right panel), and $R_{\rm long}$ (lower-left panel) of pions 
including those from $\omega$ decays (dashed curves) are a factor of 2 
to 3 larger than the radius parameters from a Gaussian fit to the 
three-dimensional correlation function obtained from the AMPT model 
(curves with squares) or to the experimentally measured one (open or 
filled circles). Excluding pions from $\omega$ decays reduces the source
radii (solid curves), and this brings $R_{\rm side}$ close to the
fitted one while $R_{\rm out}$ and $R_{\rm long}$ can still be a
factor of 2 larger than fitted ones.  The emission source from the
AMPT model thus deviates appreciably from a Gaussian one, as
found previously in studies at SPS using the RQMD transport model
\cite{Sullivan:wb,Hardtke:1999vf}. In this case, it will be useful
to compare the emission source from the AMPT model with that extracted
from measured two-particle correlation functions using the imaging
method \cite{Brown:imaging,Panitkin:1999yj}. 

\begin{figure}[ht]
\centerline{\epsfig{file=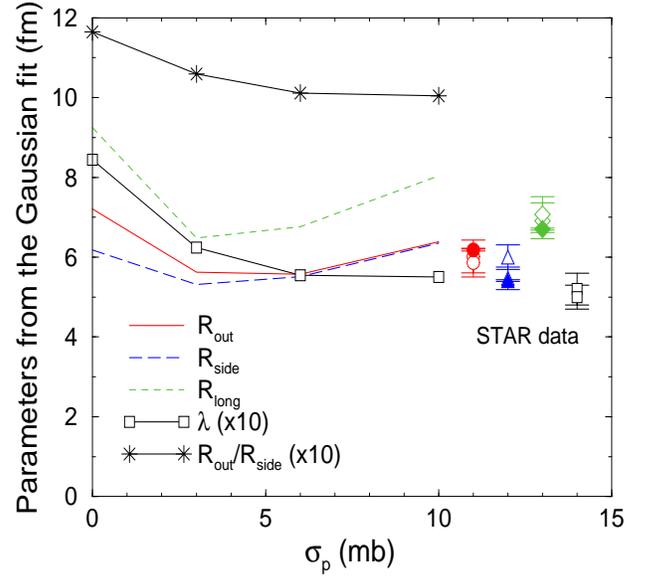,
width=3.2in,height=3.2in,angle=0}}
\caption{(Color online) 
Radii, $\lambda$ parameters and $R_{\rm out}/R_{\rm side}$
for mid-rapidity pions with $125<\pt<225$ MeV$/c$ from Gaussian fits
of the correlation function as functions of $\sigma_p$. The STAR data
at $\sqrt{s_{NN}}=130$ GeV and $200$ GeV are shown by open and filled
symbols, respectively.}
\label{r-fit}
\end{figure}

Since Eq.~(\ref{source}) gives  
\ber
{R_{\rm out}}^2&=&D_{x_{\rm out},x_{\rm out}}-2~D_{x_{\rm out},
\beta_\perp t} +D_{\beta_\perp t,\beta_\perp t},
\label{rout}
\eer
and 
${R_{\rm side}}^2=D_{x_{\rm side},x_{\rm side}}$, 
the ratio $R_{\rm out}/R_{\rm side}$ contains
information about the duration of emission and has thus been studied
extensively \cite{Rischke:1996em,Soff:2000eh,Soff:2001hc,Adler:2001zd}. 
However, the $x_{\rm out}-t$ distributions at freeze-out from the AMPT 
model show a strong positive $x_{\rm out}-t$ correlation for both
pions and kaons \cite{Lin:2002gc,Lin:2003iq}. This leads to a 
positive $D_{x_{\rm out},\beta_\perp t}$ and thus a negative
contribution to $R_{\rm out}^2$ that can be as large as the positive
duration-time term, making it difficult to extract information
about the duration of emission from the ratio 
$R_{\rm out}/R_{\rm side}$ alone. 

The dependence of fitted radii on $\sigma_p$, with $\sigma_p=0$
denoting the default AMPT model without string melting, is shown in
Fig.~\ref{r-fit} for mid-rapidity charged pions with $125<\pt<225$ MeV$/c$. 
Also shown by circles, triangles, diamonds, and squares are
the STAR data from central Au+Au collisions for mid-rapidity pions 
with $125<\pt<225$ MeV$/c$ at $\sqrt{s_{NN}}=130$ GeV (open symbols) and for 
pions with $150<\pt<250$ MeV$/c$ at $\sqrt{s_{NN}}=200$ GeV (filled symbols) 
for $R_{\rm out}$, $R_{\rm side}$, $R_{\rm long}$, and $\lambda$ 
parameter, respectively. Both the radius and the 
$\lambda$ parameters from the AMPT model with string melting 
and $\sigma_p=3-10$ mb are close to the experimental values 
\cite{Adler:2001zd}, while results from the default AMPT model
over-predict the $\lambda$ parameter. For results with string melting,
all three radius parameters are seen to increase with increasing
parton cross section $\sigma_p$ presumably as a result of the larger
source size at freeze-out, while the extracted $\lambda$ parameter
decreases gradually with increasing $\sigma_p$.

\begin{figure}[ht]
\centerline{\epsfig{file=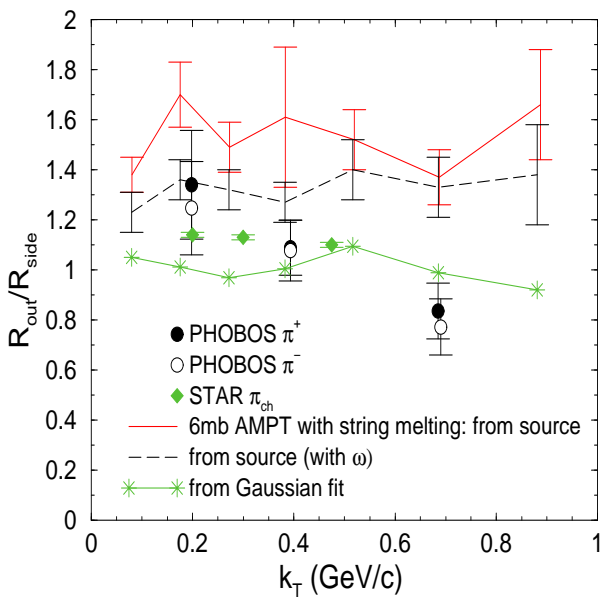,width=3.2in,
height=3.2in,angle=0}}
\caption{(Color online) 
$R_{\rm out}/R_{\rm side}$ ratios for mid-rapidity pions at
$\sqrt{s_{NN}}=200$ GeV as functions of $k_{\rm T}$.  Solid and dashed curves
are AMPT results from the emission function without and with
pions from $\omega$ decays, respectively. The curve with stars
is the AMPT result from the Gaussian fit to the two-pion correlation function.}
\label{r-ratio}
\end{figure}

Fig.~\ref{r-ratio} shows the ratio $R_{\rm out}/R_{\rm side}$ for 
mid-rapidity charged pions from AMPT with string melting and $\sigma_p=6$ mb 
for central ($b=0$ fm) Au+Au collisions at $\sqrt{s_{NN}}=200$ GeV. It is seen 
that the ratios obtained from the emission function, with solid and
dashed curves for results without and with pions from $\omega$ decays, 
have values between 1.0 and 1.7, consistent with predictions from 
the hydrodynamic model with freeze-out treated via hadronic transport
model \cite{Soff:2000eh}. However, the ratio from the Gaussian fit to 
the three-dimensional correlation function (the curve with stars) 
is much closer to 1, similar to that extracted from a Gaussian
fit to the measured correlation function 
\cite{Adler:2001zd,Adams:2003ra,Back:2004ug}. 

The AMPT model with string melting is so far the only dynamical model 
that gives a $R_{\rm out}/R_{\rm side}$ ratio close to 1 and also gives 
roughly the correct magnitude for the fitted radii. Future studies
using this transport model on the $x-t$ correlation term 
\cite{lin-heinz} or on azimuthal HBT in non-central heavy ion
collisions will help us to further test the AMPT model and 
understand the freeze-out dynamics. 

\section{Results at LHC}
\label{lhc}

\begin{figure}[ht]
\centerline{\epsfig{file=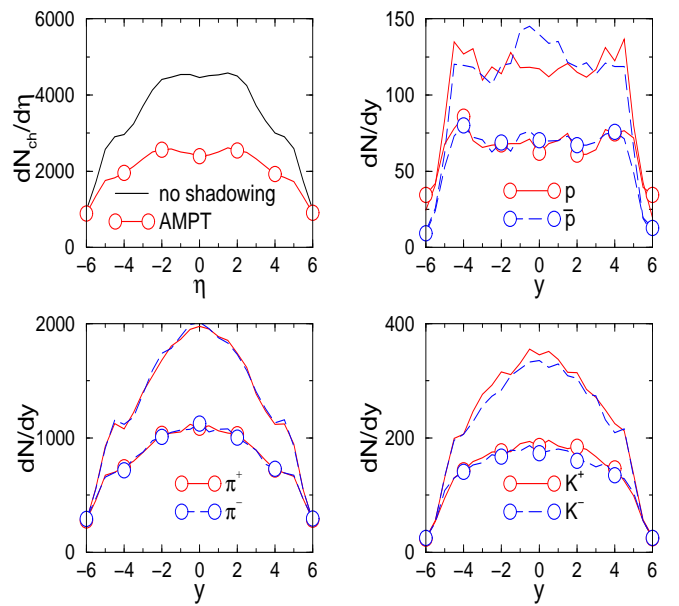,width=3.4in,height=3.2in,angle=0}}
\caption{(Color online) 
Rapidity distributions for central ($b\leq 3$ fm) Pb+Pb collisions 
at $\sqrt {s_{NN}}=5500$ GeV from the default AMPT model with (curves
with circles) and without (curves with no symbols) nuclear shadowing.}
\label{lhc-dndy}
\end{figure}

The AMPT model can also be used to study heavy ion collisions at LHC. 
Fig.~\ref{lhc-dndy} gives the charged particle pseudorapidity distribution
and the rapidity distributions of charged pions, kaons, protons, and 
antiprotons with and without nuclear shadowing in central 
($b\le 3$ fm) Pb+Pb collisions at $\sqrt{s_{NN}}=5.5$ TeV
from the default AMPT model. The distributions are significantly wider and
higher than corresponding distributions at RHIC. At mid-rapidity,
the distributions without shadowing are higher than corresponding
ones with shadowing by about 80\%. The highest value at mid-rapidity
is about 4500, well within the LHC detector limit of 7000 particles
per unit rapidity. The mid-rapidity density with nuclear shadowing is
about 2500, more than a factor of three higher than that at RHIC. 
It is higher than the logarithmic extrapolation from lower energy data
but lower than the saturation model prediction of about 3500 
\cite{Schutz:2004ji}. The charged hadron pseudorapidity distribution
shows a clear plateau structure which is very different from 
predictions from saturation models 
\cite{Armesto:2004ud,Kharzeev:2004if}. The proton and antiproton
rapidity distributions are close to each other and are almost
flat. This is different from the proton and antiproton
distributions at RHIC where protons clearly dominate at large
rapidities.  Note that the cut-off time $t_{cut}$ for hadron
cascade has been chosen as 200 fm$/c$ at LHC instead of the default
30 fm$/c$ due to the longer lifetime of the formed matter and the
larger rapidity width in heavy ion collisions at LHC.

\begin{figure}[ht]
\centerline{\epsfig{file=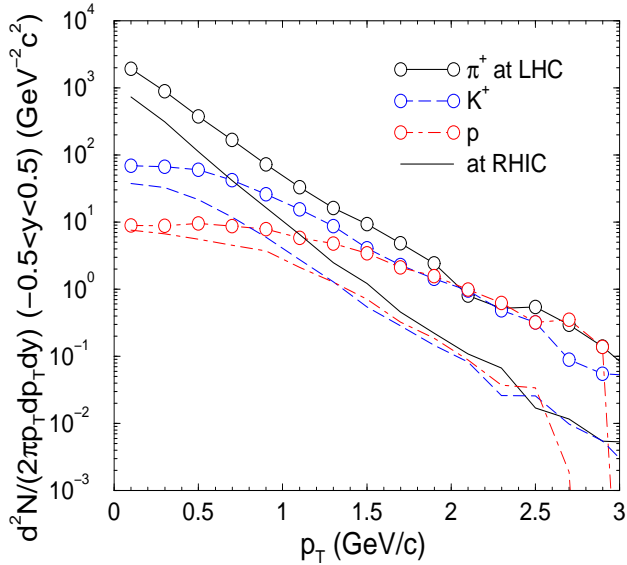,width=3.2in,height=3.2in,angle=0}}
\caption{(Color online) 
Transverse momentum spectra of pions, kaons, and protons
from the default AMPT model for central ($b\leq 3$ fm) Pb+Pb collisions 
at $\sqrt{s_{NN}}=5500$ GeV (curves with circles) 
and for central ($b\leq 3$ fm)
Au+Au collisions at $\sqrt{s_{NN}}=200$ GeV (curves with no symbols).}
\label{lhc-rhic-pt}
\end{figure}

\begin{figure}[ht]
\centerline{\epsfig{file=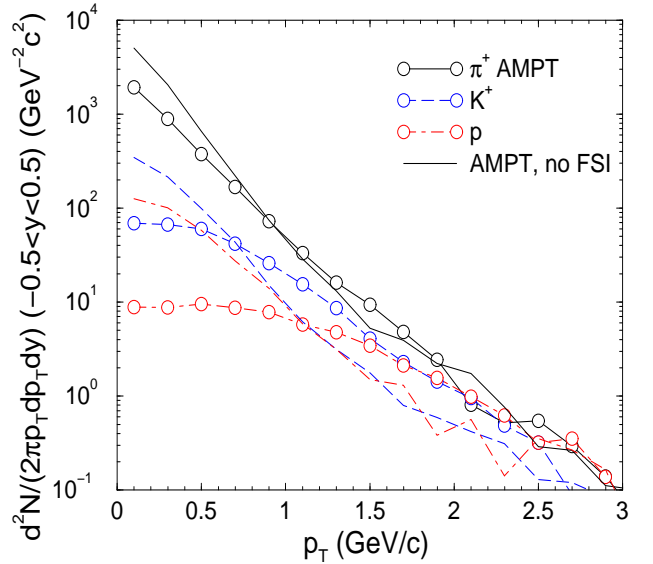,width=3.2in,height=3.2in,angle=0}}
\caption{(Color online) 
Transverse momentum spectra of pions, kaons, and protons
for central ($b\leq 3$ fm) Pb+Pb collisions at $\sqrt{s_{NN}}=5500$ GeV 
from the default 
AMPT model with (curves with circles) and without (curves with no
symbols) final-state interactions.}
\label{lhc-pt}
\end{figure}

Transverse momentum spectra at LHC are shown in Fig.~\ref{lhc-rhic-pt}. 
It is seen that the inverse slope parameters, particularly for kaons 
and protons with transverse momenta below 0.5 GeV/$c$ and 1 GeV/$c$, 
respectively, are larger than at RHIC as a result of stronger transverse 
flows. Similar to that observed at RHIC, the proton spectrum is below 
that of pions at low transverse momenta, but they become comparable 
at about 2 GeV/$c$. As in heavy ion collisions at SPS (Fig.~\ref{na44-mt}) 
and RHIC (Fig.~\ref{e200-pt-fsi}), the strong transverse flow is due to
final-state interactions, and this is shown in Fig.~\ref{lhc-pt},
where the pion, kaon, and proton spectra at LHC obtained from the AMPT
model with and without final-state interactions are compared.

\begin{figure}[ht]
\centerline{\epsfig{file=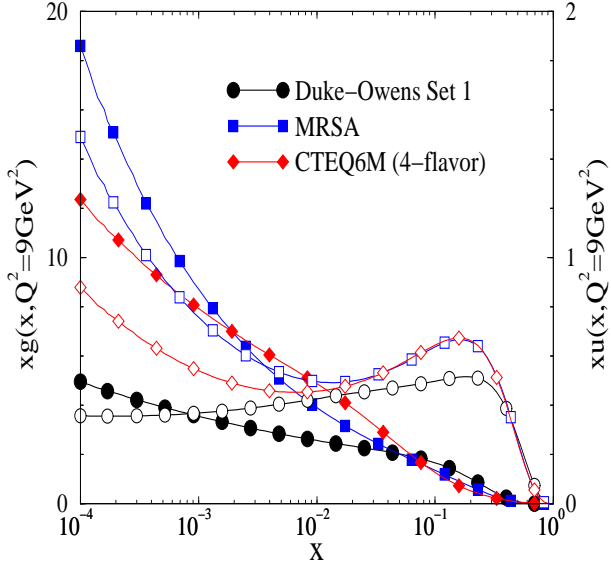,width=3.2in,
height=3.2in,angle=0}}
\caption{(Color online) 
Gluon (curves with filled symbols) and u-quark (curves with 
open symbols) distribution functions (multiplied by $x$) at $Q^2=9$ GeV$^2$ 
from three different parameterizations.}
\label{pdf}
\end{figure}

Because of the uncertainty in the parton distribution functions at 
small-$x$ in heavy nuclei, the above results on LHC have large uncertainties.
Currently, the initial condition of the AMPT model is obtained 
from the HIJING model with minijets and strings. The parton distribution 
functions used in the HIJING model are the Duke-Owens set 1
\cite{Duke:1983gd}, which are quite old.  For example, Fig.~\ref{pdf} 
shows the gluon and u-quark distribution functions at $Q^2=9$ GeV$^2$ 
from the Duke-Owens set 1 (circles), MRSA \cite{Martin:1994kn} (squares),
and the recent 4-flavor CTEQ6M \cite{Kretzer:2003it} (diamonds) 
parameterizations. It is seen that the Duke-Owens set 1 parameterization 
has far fewer partons at small-$x$, e.g., when $x < 0.01$, 
than the other two more recent parameterizations. 
At the top RHIC energy, a pair of 2 GeV back-to-back minijets 
at mid-rapidity corresponds to $x \sim 0.02$ for the initial partons, 
and at the LHC energy it corresponds to $x \sim 0.00073$. Thus 
small-$x$ partons play much more important roles at LHC than at RHIC,  
and the AMPT model will have a much larger uncertainty at LHC 
due to the underestimate of small-$x$ partons from the Duke-Owens set 1. 
An update of the parton distribution functions for the HIJING model
has been done \cite{Li:2001xa} where the Gluck-Reya-Vogt 
parton distribution functions \cite{Gluck:1994uf} have been implemented 
together with a new nuclear shadowing parametrization. To make predictions 
with better accuracy for LHC, we will need to update the parton 
distribution functions in a future version of the AMPT model. 

\section{discussions}
\label{discussions}

\begin{figure}[ht]
\centerline{\epsfig{file=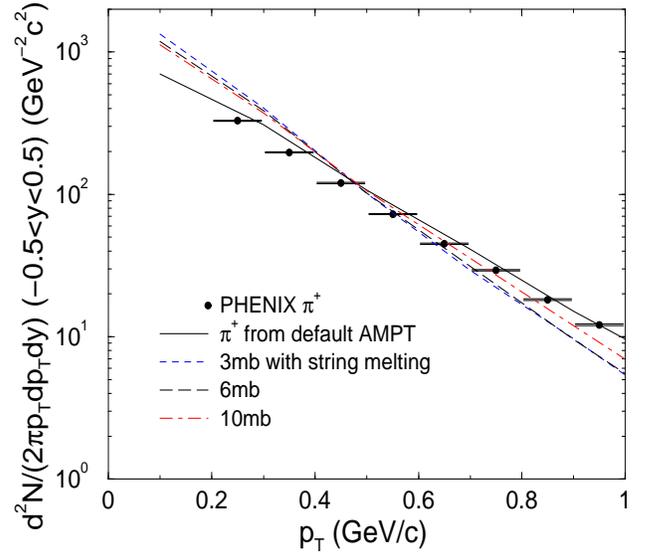,width=3.2in,
height=3.2in,angle=0}}
\caption{(Color online) 
Comparison of transverse momentum spectra of mid-rapidity pions 
from the AMPT model and the PHENIX data \protect\cite{Adler:2003cb}
for 5\% most central Au+Au collisions at $\sqrt{s_{NN}}=200$ GeV.}
\label{pt}
\end{figure}

\begin{figure}[ht]
\centerline{\epsfig{file=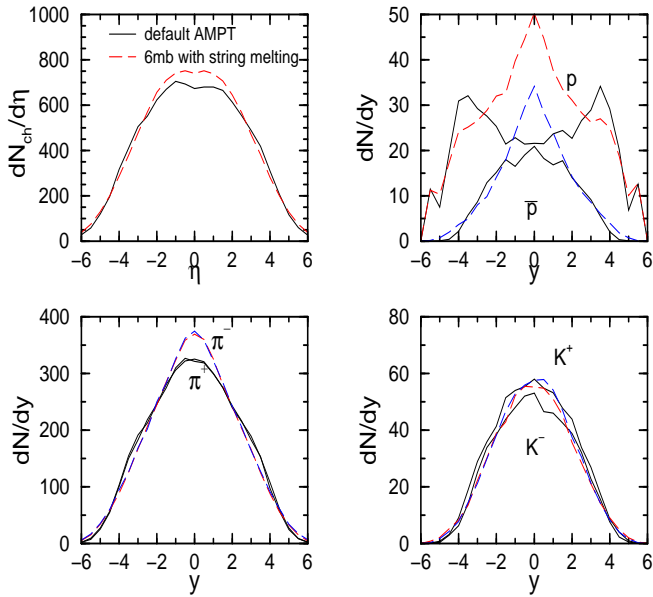,width=3.4in,
height=3.2in,angle=0}}
\caption{(Color online) Rapidity distributions of total charged particles 
(upper left panel), protons and antiprotons (upper right panel), charged 
pions (lower left panel), and charged kaons (lower right panel)
in 5\% most central Au+Au collisions at $\sqrt {s_{NN}}=200$ GeV
from the default AMPT model (solid curves) and the AMPT model 
with string melting and $\sigma_p=6$ mb (dashed curves).}
\label{dndy-sm}
\end{figure}

Two versions of the AMPT model have been used in the present study. 
The default AMPT model (current version 1.11), which includes only
minijet partons in the parton cascade and uses the Lund string model for 
hadronization, is found to give a reasonable description of hadron
rapidity distributions and transverse momentum spectra observed in
heavy ion collisions at both SPS and RHIC. It under-predicts, however,
the magnitude of the elliptic flow and also fails to reproduce the
$\lambda$ parameter of the two-pion correlation function measured at RHIC.
The latter can be described, on the other hand, by the AMPT model 
with string melting (current version 2.11) when the parton scattering
cross section is about 6 mb as shown in Sects.~\ref{v2} and \ref{hbt}.
This extended model under-predicts, however, the inverse slopes of
hadron transverse momentum spectra, and also fails to describe the 
baryon rapidity distributions as shown in Fig.~\ref{pt} and
Fig.~\ref{dndy-sm}, respectively. Since the AMPT model has not been
able to describe all experimental observables at RHIC within a single
version, further improvements are thus needed.

The initial condition of the AMPT model is obtained from the HIJING 
model with minijets and strings. Although the string melting 
mechanism \cite{Lin:2001zk,Lin:2002gc,Lin:2003iq} is introduced to 
convert the energy in initial excited strings into partons
in order to better model the partonic initial condition at high energy
densities, it is modeled by using the Lund string fragmentation to
hadrons as an intermediate process and then converting these hadrons
into their valence quarks. This is equivalent to staying at the parton 
level until strings in the Lund model have generated all the 
quark-antiquark pairs before forming hadrons. The current string melting 
can be viewed as a minimal implementation because, in the limit of 
no partonic interactions (and before hadron cascade starts), it reduces 
to HIJING results for all hadrons other than the flavor-diagonal 
mesons within the SU(2) flavor space, $\pi^0$, $\eta$, $\rho^0$ and $\omega$.
As a result, the initial partonic matter in this scenario does not 
contain gluons. Although this is unphysical as one expects gluons 
to dominate the initial stage of ultra-relativistic heavy ion collisions, 
our study of the elliptic flow \cite{Lin:2001zk} and the pion 
interferometry \cite{Lin:2002gc} depends more on the effect of 
partonic scatterings instead of the composition of the partonic 
matter. However, for dilepton production from the partonic matter, 
the flavor composition of the partonic matter is important 
\cite{Lin:1999uz}, and at present they cannot be addressed within 
the AMPT model. To extend the string melting scenario to include 
gluons in the AMPT model, we need to study the problem of 
both quark-antiquark and gluon production from a strong color 
field. Also, alternative descriptions of initial conditions, such 
as that from the parton saturation model 
\cite{McLerran:1993ni,McLerran:1993ka,Kovchegov:1997ke,Kharzeev:2000ph,Kharzeev:2001gp,Eskola:1999fc,Eskola:2000xq,Eskola:2002qz}, 
may also be used as input to the AMPT model. 

The parton cascade model ZPC in the AMPT model only includes 
leading-order two-body partonic interactions ($2$-to-$2$), while 
higher order processes ($m$-to-$n$ in general), which might
become dominant at high densities during the early stage of relativistic
heavy ion collisions \cite{Greiner:2004xp}, have not been included.  
Unfortunately, it is difficult to implement many-body interactions in
a transport model, even for the next-leading-order processes 
($gg \lra ggg$) in a parton cascade \cite{Bass:1999zq}. 
Also, we currently treat the screening mass $\mu$ as a parameter 
for changing the parton scattering cross sections.  
In principle, it should be evaluated dynamically from the effective 
temperature of the evolving parton system, leading thus to 
medium-dependent parton scattering cross sections.

Furthermore, treating interactions at a distance as in the cascade 
model used for solving the Boltzmann equations leads to possible 
violation of causality when the mean free path of partons is shorter 
than their interaction length, i.e., 
\ber
\lambda=\frac{1}{\rho \sigma} < \sqrt \frac{\sigma}{\pi}~~~ 
{\rm or~~} \frac {\rho^2 \sigma^3}{\pi}>1.
\label{pcmratio}
\eer
If we take the parton density as $\rho=10$/fm$^3$ at RHIC and take the 
parton scattering cross section as $\sigma=3$ mb, then we have
$\rho^2 \sigma^3/\pi=0.86$ and thus expect that the usual method 
used in the cascade model for solving the Boltzmann equation may 
not be very accurate. However, the Boltzmann equation with only two-body
scatterings is invariant under the transformation
\ber
f \rightarrow f ~l {\rm ~~ and ~~} \sigma \rightarrow \frac {\sigma}{l}, 
\eer
but the value of $\rho^2 \sigma^3/\pi$ is reduced by a factor of $l$. With
sufficiently large $l$, the so-called parton subdivision can then overcome
the causality problem in the cascade method  
\cite{pang,Zhang:1997ej,Zhang:1998tj,Molnar:2000jh,Molnar:v2,Cheng:2001dz}.
Although parton subdivision has been implemented in ZPC, 
it remains to be implemented in the AMPT model mainly due to the 
complication in the hadronization process which converts partons into 
hadrons following the parton cascade. How the causality violation 
in the AMPT model affects the final observables in heavy ion 
collisions \cite{Molnar:v2} is yet to be studied.

In the hadronization process in the default AMPT model, a minijet 
recombines with its parent string even though the minijet has gone 
through partonic interactions and thus has changed its color charge. 
The HIJING model also takes a similar approach in the jet quenching process 
because a gluon is still being associated with the same parent string 
after it goes through inelastic collisions. Our current prescription 
guarantees that fragmentation can successfully proceed for all 
strings after parton interactions, and it also enables the default 
AMPT model to reduce to HIJING results in the limit of no partonic 
(and hadronic) interactions. Although the freeze-out positions and 
times of minijet partons are averaged in a string before its fragmentation, 
the current treatment of color configurations certainly needs to be improved, 
where new color-singlet strings over the whole volume need to be 
reconstructed after partonic interactions.

To describe charged particle multiplicities in heavy ion collisions 
at SPS, we have changed the values of the $a$ and $b$ parameters 
in the Lund symmetric fragmentation function given by 
Eq.~(\ref{lundsf}) based on the possibility that these parameters 
could be modified in the dense matter formed in heavy ion collisions. 
However, the AMPT model needs the default $a$ and $b$ values, i.e.,  
the values in the HIJING model, in order to reproduce the charged particle 
multiplicities in $pp$ and $p\bar p$ collisions.  The $a$ and $b$ 
values in the AMPT model is thus expected to depend on the atomic 
weight of colliding nuclei and the centrality of their collisions.  
Although the AMPT results on $d+Au$ collisions using default $a$ 
and $b$ values \cite{Lin:2003ah} agree reasonably with the RHIC 
$d+Au$ data \cite{Back:2003hx,Adams:2004dv,Simon:2004di}, 
the detailed dependence of $a$ and $b$ parameters on the system size 
has not been studied. 

At present, phase transition in the AMPT model with string melting  
is modeled by a simple quark coalescence using the current quark masses.
The failure of the proton rapidity distribution in this model, shown
in Fig.~\ref{dndy-sm}, is related to this assumption as the proton
mass is not generated dynamically but is given to the coalescing
system of three light quarks. In the case that the invariant mass of
the three quarks is small, the resulting proton with its physical mass
tends to overpopulate at mid-rapidity. This problem may be avoided if 
protons are formed from coalescence of quarks with constituent quark
masses. However, the current string melting mechanism would fail to convert 
strings into partons since the pion, as a Goldstone boson, cannot be 
decomposed into a quark and an antiquark with constituent quark masses. 
A more consistent method is perhaps to use density or temperature
dependent quark masses, so that they correspond to current masses at
high temperature and constituent masses at low temperature near the
phase transition, where a scalar field is responsible for the changing 
masses. This method also has the advantage that it can qualitatively 
describe the equation of state of the QGP \cite{Danielewicz:1998vz}. 
Also, the current coalescence model could have problems with entropy, 
because quark coalescence reduces the number of particles by a 
factor of two to three, although entropy also depends on the degeneracy 
and mass of produced hadrons. We have studied the entropy problem 
in a schematic model using a thermalized QGP with the same parameters 
as given in Ref.~\cite{Greco:2004ex}. Because of productions of massive 
resonances such as $\rho, \omega$, and $\Delta$, the total entropy is 
reduced by about 15\% when partons are converted to hadrons 
via coalescence, and this is mainly due to a similar violation in energy. 
To eliminate entropy violation thus requires a proper treatment
of energy conservation. This requires the inclusion of field energy 
in both QGP and hadronic matter, which is also necessary for 
properly describing the equation of state. Furthermore, although 
the spatial correlation in the quark coalescence based on 
the closest neighbors leads to momentum correlation among coalescing 
partons in the presence of collective flow, explicit momentum 
correlation is not considered in our quark coalescence model, 
and random momentum distributions of coalescing quarks tend to 
give the artificial peak at mid-rapidity shown in Fig.~\ref{dndy-sm}. 
The quark coalescence model can be improved by following the method
used recently in studying hadron production from the quark-gluon
plasma, which has been shown to describe satisfactorily the observed 
quark number scaling of hadron elliptic flows and large baryon to pion
ratio at intermediate transverse momenta
\cite{Lin:2002rw,Voloshin:2002wa,Molnar:2003ff,Fries:2003vb,Greco:2003xt,Lin:2003jy,Greco:2003vf,Greco:2003mm,Fries:2003kq}. 
Also, statistical models, which generate hadronic states according to 
their statistical weights, provides an alternative method 
to describe the parton-to-hadron phase transition. 

Many of the hadronic cross sections used in the hadronic stage of the
AMPT model have not been studied theoretically in detail or well
constrained by experimental data, and they may be important for some
observables. For example, from the study at SPS energies with the
default AMPT model, we have found that the explicit inclusion of $K^*$ 
mesons and baryon-antibaryon production from two-meson states are
important for strangeness and antiproton production, respectively. 
In the AMPT model with string melting, the phase transition
happens later than in the default AMPT model due to the larger energy
and lifetime of the partonic system, hadronic effects are less
important than in the default AMPT model. Even in this case, it is
essential to know the cross sections of $\eta$ meson interactions with
other hadrons as they  may affect the final $\eta$ yield and consequently 
the height (or the $\lambda$ parameter) of the two-pion correlation function. 
Effects due to the uncertainties of these hadronic input parameters 
on final observables at RHIC will be studied in the future.

\section{summary}
\label{summary}

To study high energy heavy ion collisions at SPS, RHIC and even higher
energies such as at the LHC, a multi-phase transport (AMPT) model has
been developed. It consists mainly of four components: the HIJING
model to convert the initial incident energy to the production of hard 
minijet partons and soft strings, with excited strings further converting
to partons in the AMPT model with string melting; Zhang's parton 
cascade (ZPC) to model the interactions among partons; the Lund string 
fragmentation as implemented in JETSET/PYTHIA to convert the excited
strings to hadrons in the default model or a simple quark coalescence
model to convert partons into hadrons in the case of string melting;
and the extended relativistic transport (ART) model for describing 
interactions among hadrons. In this paper, we have described in detail the 
physics input in each component and how the different components are 
combined to give a comprehensive description of the dynamics of 
relativistic heavy ion collisions.  We have used this model to study 
various observables in heavy ion collisions and to address the
relative importance of partonic and hadronic effects on these
observables. In particular, the AMPT model has been used to study 
the rapidity distributions of particles such as pions, kaons, protons 
and antiprotons, their transverse momentum spectra, the elliptic flow,
and the interferometry of two identical mesons. We find that the
default AMPT model (current version 1.11) gives a reasonable
description of rapidity distributions and transverse momentum spectra,
while the AMPT model with string melting (current version 2.11)
describes both the magnitude of the elliptic flow at mid-rapidity 
and the pion correlation function with a parton cross section of about 6 mb.  

High energy heavy ion collisions is a complex process involving the initial 
conditions, the interactions of initially produced partons and of 
later hadronic matter as well as the transition between these two 
phases of matter. The AMPT model is an attempt to incorporate these 
different physics as much as we can at present. Since there are many 
uncertainties in the AMPT model, the model is currently more a 
simulation tool rather than a finalized code. Nevertheless, this model 
provides the possibility to study the dependence of various 
observables on these physical effects. For example, we have found 
that the elliptic flows are sensitive to early parton dynamics, and 
the HBT interferometry is instead affected by the complicated late 
hadron freeze-out dynamics. The AMPT model can be extended, e.g., 
to include hydrodynamic evolution at the early stage when local 
thermalization is likely, in order to conveniently study the 
equation of state of the partonic matter.  Experiments at RHIC 
such as d+A can also help to reduce the uncertainties in the physical 
input such as the parton distribution functions in heavy nuclei. 
With continuing efforts in both theoretical and experimental heavy 
ion physics, we hope that this multi-phase transport model will 
eventually incorporate the essential elements of the underlying 
theory of QCD to provide a reliable description of different 
observables in heavy ion collisions within one coherent
picture, and help us to learn from relativistic heavy ion collisions
the properties of the quark-gluon plasma formed during the early stage
of the collisions.

\begin{acknowledgments}
We thank L.W. Chen, V. Greco, M. Gyulassy, U. Heinz, H. Huang, 
D. Molnar, M. Murray, and N. Xu for useful discussions. Z.W.L. and 
B.Z. thank the Department of Energy Institute for Nuclear Theory at 
the University of Washington for hospitality during the INT-03-1 
program ``The first three years of heavy ion physics at RHIC" where 
part of the work was done. We also thank the Parallel Distributed
System Facility at the National Energy Research Scientific Computer
Center for providing computer resources. This work was supported by 
the U.S. National Science Foundation under Grant Nos. PHY-0098805 
(Z.W.L., S.P., and C.M.K.), PHY-0457265 (C.M.K.), PHY-0354572 
(B.A.L.), and PHY-0140046 (B.Z.), the U.S. Department of Energy under 
Grant No. DE-FG02-01ER41190 (Z.W.L.), the Welch Foundation under Grant 
No. A-1358 (Z.W.L., S.P., and C.M.K.).
\end{acknowledgments}

\appendix* 
\section{AMPT Users' Guide}
The default AMPT model (current version 1.11) and the AMPT model with
string melting (current version 2.11) both use an initialization file 
`input.ampt'. The analysis directory `ana/' contains the resulting
data files. The final particle record file is `ana/ampt.dat'.
The version number of AMPT is written to both `ana/version'
and `nohup.out' files.  The AMPT source code has been tested for both 
f77 and pgf77 compilers on the UNIX, Linux, and OSF1 operating systems.

To run the AMPT program, one needs to:
\begin{enumerate}
\item set the initial parameters in `input.ampt'. If one prefers to
use run-time random number seed, set `ihjsed=11', In this way, every
run is different even with the same `input.ampt' file.
\item type `\texttt{sh exec \&}' to compile and run the executable `ampt'
with some general information written in `nohup.out'.
\end{enumerate}

Key initial parameters in `input.ampt' are:
\begin{description}
\item EFRM: $\sqrt{s_{NN}}$ in GeV, e.g. 200 for the maximum RHIC energy.
\item NEVNT: the total number of events.
\item BMIN, BMAX: the minimum and maximum impact parameters (in fm) 
   for all events with BMAX having an upper limit of HIPR1(34)+HIPR1(35) 
   (=19.87 fm for d+Au collisions and 25.60 fm for Au+Au collisions). 
\item ISOFT: choice of parton-hadron conversion scenario.\\
   =1: default AMPT model (version $1.x$);\\
   =4: the AMPT model with string melting (version $2.y$).\\
   Note that values of 2, 3, and 5 have never been used for publications.
   They are tests of other string melting scenarios:\\
   =2: a string is decomposed into q+qq+minijet partons instead of 
       using the Lund fragmentation;\\
   =3: a baryon is decomposed into q+qq instead of 3 quarks;\\
   =5: same as 4 but partons freeze out according to
      local energy density.
\item NTMAX: the number of time-steps for hadron cascade, default(D)=150.
   Note that NTMAX=3 effectively turns off hadron cascade, 
   and a larger than the default value is usually necessary 
   for observables at large rapidity or large pseudorapidity.
   We use NTMAX=1000 for HBT studies in central Au+Au
   collisions due to the need for the space-time information of last 
   interactions and for LHC calculations due to the longer lifetime 
   of the formed matter. 
\item DT: value of the time-step (in fm$/c$) for hadron cascade, D=0.2.
   Note that $t_{cut}=\text{NTMAX} \times \text{DT}$ is the 
   termination time of hadron cascade. 
\item PARJ(41): parameter $a$ in the Lund symmetric fragmentation function. 
\item PARJ(42): parameter $b$ in the Lund symmetric fragmentation function 
   (in GeV$^{-2}$). Note that we use default value in HIJING ($a=0.5$ and 
   $b=0.9$) for d+Au collisions, 
   and $a=2.2$ and $b=0.5$ for collisions of heavy nuclei.
\item flag for popcorn mechanism: D=1(Yes) 
   turns on the popcorn mechanism. In general, it increases baryon stopping.
\item PARJ(5): controls $\mathrm{BM\bar{B}}$ vs. $\mathrm{B\bar{B}}$ 
   in the popcorn mechanism, D=1.0. 
\item shadowing flag: D=1(Yes) turns on nuclear shadowing. 
\item quenching flag: D=0(No) turns off jet quenching 
   since the parton cascade ZPC simulates final-state effects. 
\item p0 cutoff: D=2.0 (in GeV$/c$) for $p_0$ in HIJING for minijet 
   production. 
\item parton screening mass: controls the parton cross section, 
   D=3.2264 (in fm$^{-1}$).
   Its square is inversely proportional to the parton cross
   section. Use D=3.2264 for 3mb, and 2.2814 for 6mb.
\item ihjsed: choice of the random number seed, D=0.\\
   =0: take the `Ran Seed for HIJING' in `input.ampt'
   and disregard the random value generated in the file `exec'.\\
   =11: take the HIJING random seed at runtime
   from the file `exec', with the seed written
   in `nohup.out' and `ana/version'.
\item Ran Seed for HIJING: random number seed for HIJING when ihjsed=0.
\item Kshort decay flag: depends on the experimental correction procedure, 
   D=0 turns off Kshort decays after the hadron cascade.
   Note that decays of the following resonances and their
   antiparticles are always included: 
   $\rho$, $\omega$, $\eta$, K*, $\phi$, $\Delta$, N*(1440), N*(1535),
   $\Sigma^0$ (in order to include its feed down to $\Lambda$). 
\item optional OSCAR output: if set to 1, outputs in OSCAR1997A format
   \cite{oscar} are written in `ana/parton.oscar' and `ana/hadron.oscar'. 
\item dpcoal: parton coalescence distance in momentum space (in GeV$/c$).
\item drcoal: parton coalescence distance in coordinate space (in fm).
   dpcoal, drcoal both have D=$10^6$ for nearest-neighbor coalescence 
   in the AMPT model with string melting. 
\end{description}

Key output file are:
\begin{description}
\item ana/ampt.dat: It contains particle records at hadron kinetic freeze-out, 
  i.e., at the last interaction. For each event, the first line gives: 
        event number, test number(=1), number of particles in the event,
        impact parameter, total number of participant nucleons in projectile,
        total number of participant nucleons in target, number of participant
        nucleons in projectile due to elastic collisions, number of
        participant nucleons in projectile due to inelastic collisions,
        and corresponding numbers in target.
        Note that participant nucleon numbers include nucleons participating
        in both elastic and inelastic collisions.
      Each of the following lines gives:
PYTHIA particle ID number, three-momentum($p_x$,$p_y$,$p_z$), mass, and 
space-time coordinates($x$,$y$,$z$,$t$) of one final particle at freeze-out.
Note that momenta are in units of GeV$/c$, mass in GeV$/c^2$,
space in fm, and time in fm$/c$.
If a particle comes from the decay of a resonance which still exists
at the termination time of hadron cascade, then its space-time
corresponds to the decay point of the parent resonance.
Also note that the x-axis in the AMPT program is defined as the direction along
      the impact parameter, and the z-axis is defined as the beam direction.
\item ana/zpc.dat:    similar to `ana/ampt.dat' but for partons.
          The first line of each event gives:
        event number, number of partons in the event, impact parameter,
         number of participant nucleons in projectile due to elastic
         collisions, number of participant nucleons in projectile due to
         inelastic collisions, and corresponding numbers in target.
      Each of the following lines gives:
PYTHIA particle ID number, three-momentum($p_x$,$p_y$,$p_z$), mass, and 
space-time coordinates($x$,$y$,$z$,$t$) of one final parton at freeze-out.
\end{description}


\begin{thebibliography}{999}
\bibitem{smu2} 
We have checked numerically that the average $\sqrt s$ in AMPT is 
around 500 to 600 MeV, which is comparable to the default value of 
$\mu=3$ fm$^{-1}$ but not larger. 
Eq.~(\ref{sgg}) thus might be oversimplified.
\bibitem{ablund}
The general Lund flavor-dependent symmetric fragmentation function 
involves two $a$ parameters which are different for quark and 
diquark fragmentation/production, while the $b$ parameter is universal. 
As in the default PYTHIA model, the $a$ parameter for diquarks is larger 
than that for quarks by $0.5$ in the AMPT model. 
Eq.~(\ref{lundsf}) is the simplified expression, where $a$ 
corresponds to quark production in a $q\bar q$ string.
\bibitem{ftime} 
For central Pb+Pb collisions at the SPS (see Figs.~\ref{na49-dndy-hjab} 
and \ref{na49-dndy-newab}), we find that with the formation time
increased from 0.7 fm/c to 1.2 fm/c in the default AMPT model the
rapidity distributions of negative hadrons and pions are slightly 
higher (by about 3\% on the average), protons change within about 
3\%, kaons are lower by about 5\%, and antiprotons are lower by about 10\%. 
The decrease in kaons and increase in pions are due to fewer hadronic 
interactions which produce strangeness, and the decrease in antiprotons 
is due to a smaller diffusion of baryons into the mid-rapidity region.
\bibitem{energy}
We have checked numerically that the violation of total energy in any event 
with string melting is between -0.4\protect \% and 1.7\protect \% 
for 5000 minimum-bias Au+Au events at $\sqrt{s_{NN}}=130$ GeV.
However, the violation in transverse energy could be much larger.
For example, take $\delta y \sim 6$, $\langle E_T \rangle \sim 0.5$ GeV 
and $dN_{total}/dy\sim 1350$ for central Au+Au collisions at 
$\sqrt{s_{NN}}=200$ GeV  
and attribute all the violation in total energy to the transverse direction 
for a worst-case estimate,  
then an 1\protect \% violation (or 400 GeV) in total 
energy translates to 67 GeV violation in transverse energy at mid-rapidity, 
a violation of up to 10\protect \%.
\bibitem{amptjpsi}
The current version of the AMPT model (v1.11 and v2.11) 
does not include the option of treating $J/\psi$ dynamics.
\end{thebibliography}
\end{document}